\documentclass[a4paper]{amsart}

\RequirePackage{amsmath}
\RequirePackage{bm}
\RequirePackage{amssymb}
\RequirePackage{upref}
\RequirePackage{amsthm}
\RequirePackage{enumerate}
\RequirePackage{pb-diagram}
\RequirePackage{amsfonts}
\RequirePackage[mathscr]{eucal}
\RequirePackage{verbatim}
\RequirePackage{xr}
\RequirePackage{graphicx}
\usepackage{calc}
\usepackage{xspace}
\RequirePackage{color}

\newcommand{\cf}{cf.\@\xspace}
\newcommand{\resp}{resp.\@\xspace}

\newcommand{\al}{\alpha}
\newcommand{\bet}{\beta}
\newcommand{\ga}{\gamma}
\newcommand{\de}{\delta }
\newcommand{\e}{\epsilon}

\newcommand{\f}{\varphi}
\newcommand{\h}{\eta}

\newcommand{\ka}{\kappa}
\newcommand{\lam}{\lambda}
\newcommand{\m}{\mu}
\newcommand{\n}{\nu}

\newcommand{\s}{\sigma}
\newcommand{\x}{\xi}

\newcommand{\D}{\varDelta}

\newcommand{\Lam}{\varLambda}
\newcommand{\Om}{\varOmega}

\newcommand\whc[1]{\underset{#1}\rightharpoondown} 

\newcommand{\di}[1]{#1\nobreakdash-\hspace{0pt}dimensional}

\newcommand{\fv}[2]{#1\hspace{0pt}_{|_{#2}}}

\newcommand{\so}{{\mc S_0}}

\newcommand{\const}{\tup{const}}
\newcommand{\ndash}{\nobreakdash--}

\newcommand{\msp[1]}[1]{\mspace{#1mu}}

\newcommand{\R}[1][n+1]{{\protect\mathbb R}^{#1}}

\newcommand{\Cc}{{\protect\mathbb C}}

\newcommand{\N}{{\protect\mathbb N}}

\newcommand{\eR}{\stackrel{\lower1ex \hbox{\rule{6.5pt}{0.5pt}}}{\msp[3]\R[]}}
\newcommand{\eN}{\stackrel{\lower1ex \hbox{\rule{6.5pt}{0.5pt}}}{\msp[1]\N}}
\newcommand{\eO}{\stackrel{\lower1ex
\hbox{\rule{6pt}{0.5pt}}}{\msc O}}

\DeclareMathOperator{\graph}{graph}

\DeclareMathOperator{\grad}{grad}

\newcommand\im{\implies}
\newcommand\ra{\rightarrow}

\newcommand\hra{\hookrightarrow}

\newcommand\pa{\partial}
\newcommand\pde[2]{\frac {\partial#1}{\partial#2}}
 
\newcommand\df[2]{\frac {d#1}{d#2}}

\newcommand{\un}{\infty}
\newcommand{\A}{\forall}

\newcommand{\set}[2]{\{\,#1\colon #2\,\}}
\newcommand{\uu}{\cup}

\newcommand{\uuu}{\bigcup}

\newcommand{\uud}{ \stackrel{\lower 1ex \hbox {.}}{\uu}}
\newcommand{\uuud}[1]{ \stackrel{\lower 1ex \hbox {.}}{\uuu_{#1}}}
\newcommand\su{\subset}

\newcommand\eS{\emptyset}
\newcommand{\sminus}[1][28]{\raise 0.#1ex\hbox{$\scriptstyle\setminus$}}

\newcommand{\wt}{\widetilde}

\newcommand{\wed}{\wedge}

\newcommand{\abs}[1]{\lvert#1\rvert}

\newcommand{\norm}[1]{\lVert#1\rVert}

\newcommand{\spd}[2]{\protect\langle #1,#2\protect\rangle}

\newcommand\ch[3]{\varGamma_{#1#2}^#3}
\newcommand\cha[3]{{\bar\varGamma}_{#1#2}^#3}

\newcommand{\riem}[4]{R_{#1#2#3#4}}
\newcommand{\riema}[4]{{\bar R}_{#1#2#3#4}}

\newcommand{\tit}{\textit}

\newcommand{\tup}{\textup}

\newcommand{\mc}{\protect\mathcal}
\newcommand{\msc}{\protect\mathscr}

\providecommand{\bysame}{\makebox[3em]{\hrulefill}\thinspace}

\newcommand{\ci}{\cite}

\newcommand{\cq}[1]{\glqq{#1}\grqq\,}

\newcommand{\bt}{\begin{thm}}
\newcommand{\bl}{\begin{lem}}
\newcommand{\bc}{\begin{cor}}
\newcommand{\bd}{\begin{definition}}
\newcommand{\bpp}{\begin{prop}}
\newcommand{\br}{\begin{rem}}
\newcommand{\bn}{\begin{note}}
\newcommand{\be}{\begin{ex}}
\newcommand{\bes}{\begin{exs}}
\newcommand{\bb}{\begin{example}}
\newcommand{\bbs}{\begin{examples}}
\newcommand{\ba}{\begin{axiom}}
\newcommand{\bas}{\begin{assumption}}

\newcommand{\et}{\end{thm}}
\newcommand{\el}{\end{lem}}
\newcommand{\ec}{\end{cor}}
\newcommand{\ed}{\end{definition}}
\newcommand{\epp}{\end{prop}}
\newcommand{\er}{\end{rem}}
\newcommand{\en}{\end{note}}
\newcommand{\ee}{\end{ex}}
\newcommand{\ees}{\end{exs}}
\newcommand{\eb}{\end{example}}
\newcommand{\ebs}{\end{examples}}
\newcommand{\ea}{\end{axiom}}
\newcommand{\eas}{\end{assumption}}

\newcommand{\bp}{\begin{proof}}
\newcommand{\ep}{\end{proof}}
\newcommand{\eps}{\renewcommand{\qed}{}\end{proof}}

\newcommand{\bal}{\begin{align}}

\newcommand{\bi}[1][1.]{\begin{enumerate}[\upshape #1]}
\newcommand{\bia}[1][(1)]{\begin{enumerate}[\upshape #1]}
\newcommand{\bin}[1][1]{\begin{enumerate}[\upshape\bfseries #1]}
\newcommand{\bir}[1][(i)]{\begin{enumerate}[\upshape #1]}
\newcommand{\bic}[1][(i)]{\begin{enumerate}[\upshape\hspace{2\cma}#1]}
\newcommand{\bis}[2][1.]{\begin{enumerate}[\upshape\hspace{#2\parindent}#1]}
\newcommand{\ei}{\end{enumerate}}

\newcommand\ndots{\raise 0.47ex \hbox {,}\hskip0.06em\cdots %
     \raise 0.47ex \hbox {,}\hskip0.06em} 

\newcommand{\q}{\quad}
\newcommand{\qq}{\qquad}

\newcommand{\hp}{\hphantom}

\newcommand\nd{\noindent}

\newskip\Csmallskipamount                                                
\Csmallskipamount=\smallskipamount
\newskip\Cmedskipamount
\Cmedskipamount=\medskipamount
\newskip\Cbigskipamount
\Cbigskipamount=\bigskipamount

\newcommand\cvs{\vspace\Csmallskipamount}   
\newcommand\cvm{\vspace\Cmedskipamount}

\newskip\csa
\csa=\smallskipamount

\newskip\cma
\cma=\medskipamount

\newskip\cba
\cba=\bigskipamount

\newdimen\spt
\newcommand\citem{\cvs\advance\itemno by
1{(\romannumeral\the\itemno})\hskip3pt}
\newcommand{\bitem}{\cvm\nd\advance\itemno by
1{\bf\the\itemno}\hspace{\cma}}

\newcount\itemno
\newcommand{\las}[1]{\label{S:#1}}
\newcommand{\lass}[1]{\label{SS:#1}}
\newcommand{\lae}[1]{\label{E:#1}}
\newcommand{\lat}[1]{\label{T:#1}}
\newcommand{\lal}[1]{\label{L:#1}}
\newcommand{\lad}[1]{\label{D:#1}}

\newcommand{\lar}[1]{\label{R:#1}}

\newcommand{\rs}[1]{Section~\ref{S:#1}}

\newcommand{\rt}[1]{Theorem~\ref{T:#1}}
\newcommand{\rl}[1]{Lemma~\ref{L:#1}}
\newcommand{\rd}[1]{Definition~\ref{D:#1}}

\newcommand{\re}[1]{\eqref{E:#1}}
\newcommand{\frt}[1]{Theorem~\ref{T:#1} on page~\tup{\pageref{T:#1}}}
\newcommand{\frl}[1]{Lemma~\ref{L:#1} on page~\tup{\pageref{L:#1}}}
\newcommand{\frr}[1]{Remark~\ref{R:#1} on page~\tup{\pageref{R:#1}}}
\newcommand{\frd}[1]{Definition~\ref{D:#1} on page~\tup{\pageref{D:#1}}}
\newcommand{\fre}[1]{\eqref{E:#1} on page~\tup{\pageref{E:#1}}}
\newcommand{\frss}[1]{Subsection~\ref{SS:#1} on page~\tup{\pageref{SS:#1}}}

\newskip\thmskip
\thmskip=\parindent

\newskip\hsk
\newenvironment{hinw}{\labelsep=0pt\begin{list}{}{\labelsep=0pt\itemindent=0pt\labelwidth=0pt\leftmargin=\parindent\rightmargin=0pt\partopsep=\cba}%
\item\it\nopagebreak\nopagebreak}%
{\end{list}}

\newcommand\bh{\begin{hinw}}
\newcommand{\eh}{\end{hinw}}

\newtheoremstyle{normal}
  {\cba}
  {\cba}
  {}
  {\thmskip}
  {\bfseries}
  {.}
  {\hsk}
  {}

\newtheoremstyle{abschnitt}
  {\cba}
  {\cba}
  {}
  {\thmskip}
  {\bfseries}
  {.}
  {\hsk}
  {}

\newtheoremstyle{italic}
  {\cba}
  {\cba}
  {\itshape}
  {\thmskip}
  {\bfseries}
  {.}
  {\hsk}
  {}

\newtheoremstyle{aufgaben}
  {\cba}
  {\cba}
  {}
  {}
  {\normalsize\bfseries}
  {.}
  {\hsk}
  {}

\newtheoremstyle{break}
  {\cba}
  {\cba}
  {\itshape}
  {}
  {\bfseries}
  {.}
  {\newline}
  {}

\swapnumbers
\theoremstyle{italic}
\newtheorem{thm}[subsection]{Theorem}
\newtheorem{lem}[subsection]{Lemma}
\newtheorem{prop}[subsection]{Proposition}
\newtheorem{cor}[subsection]{Corollary}

\theoremstyle{normal}
\newtheorem{rem}[subsection]{Remark}
\newtheorem{definition}[subsection]{Definition}
\newtheorem{example}[subsection]{Example}
\newtheorem{examples}[subsection]{Examples}
\newtheorem{ex}[subsection]{Exercise}
\newtheorem{note}[subsection]{}
\newtheorem{axiom}[subsection]{Axiom}
\newtheorem{assumption}[subsection]{Assumption}

\theoremstyle{aufgaben}
\newtheorem{exs}[subsection]{Exercises}

\swapnumbers

\numberwithin{equation}{section}
\numberwithin{figure}{section}

\newenvironment{textequation}[1][0.8]
{\begin{equation}
\begin{aligned}
\begin{minipage}{#1\linewidth}}
{\end{minipage}
\end{aligned}
\end{equation}
\ignorespacesafterend}

\newcommand{\btext}{\begin{textequation}}
\newcommand{\etext}{\end{textequation}}

\def\hinweis{\@startsection{subsection}{2}%
 \z@{0.7\linespacing\@plus 0.5\linespacing}{0.7\linespacing}%
{\normalfont\itshape\indent}}

\usepackage[german,english]{babel}
\usepackage{graphicx}
\RequirePackage{amsmath}
\RequirePackage{bm}
\RequirePackage{amssymb}
\RequirePackage{upref}
\RequirePackage{amsthm}
\RequirePackage{enumerate}
\RequirePackage{pb-diagram}
\RequirePackage{amsfonts}
\RequirePackage[mathscr]{eucal}
\RequirePackage{verbatim}
\RequirePackage{xr}
\RequirePackage{graphicx}
\usepackage{calc}
\usepackage{xspace}

\makeatletter
\RequirePackage{color}
\newcommand{\ann}[1]{\renewcommand{\@makefnmark}{\mbox{$^{\color{red}{\@thefnmark}}$}}%
\footnote {#1}}
\makeatother

\RequirePackage{upref}
\RequirePackage{amsthm}
\RequirePackage{enumerate}
\usepackage[mathscr]{eucal}

\usepackage{xr-hyper}

\usepackage{calc}

\newlength{\oddsidemarginlength}
\newlength{\topmarginlength}

\newcounter{numberoflines}
\newcounter{tempcc}
\oddsidemargin=\oddsidemarginlength
\evensidemargin=\oddsidemargin
\topmargin=\topmarginlength
\begin{document}

\flushbottom

\mbox{}
\vskip -1cm


\title[Quantum cosmological models]{Quantum cosmological Friedman models with an initial singularity}

\author{Claus Gerhardt}
\address{Ruprecht-Karls-Universit\"at, Institut f\"ur Angewandte Mathematik,
Im Neuenheimer Feld 294, 69120 Heidelberg, Germany}
\email{gerhardt@math.uni-heidelberg.de}
\urladdr{http://www.math.uni-heidelberg.de/studinfo/gerhardt/}
\thanks{This work has been supported by the DFG}

%
\date{\today}
%


\begin{abstract}
We consider the Wheeler-DeWitt equation $H\psi=0$ in a suitable Hilbert space. It turns out that this equation has countably many solutions $\psi_i$ which can be considered as eigenfunctions of a Hamilton operator implicitly defined by $H$. We consider two models, a bounded one, $0<r<r_0$, and an unbounded, $0<r<\un$, which represent different eigenvalue problems. In the bounded model we look for eigenvalues $\Lam_i$, where the $\Lam_i$ are the values of the cosmological constant which we used in the Einstein-Hilbert functional, and in the unbounded model the eigenvalues are given by $(-\Lam_i)^{-\frac {n-1}{n}}$, where $\Lam_i<0$. Notice that $r$ is the symbol for the scale factor, usually denoted by $a$, or a power of it.

The $\psi_i$ form a basis of the underlying Hilbert space. We prove furthermore that the implicitly defined Hamilton operator is selfadjoint and that the solutions of the corresponding Schr\"odinger equation satisfy the Wheeler-DeWitt equation, if the initial values are  superpositions of eigenstates.

All solutions have an initial singularity in $r=0$. Under certain circumstances a smooth transition from big crunch to big bang is possible.
\end{abstract}

\maketitle

\tableofcontents

\setcounter{section}{0}
\section{Introduction}\las{1}
Arnowit, Deser and Misner showed in their celebrated paper \cite{adm:old} how the Lagrangian formulation of general relativity can be expressed in a way allowing to apply the Legendre transformation to obtain a Hamiltonian formulation. Since the Lagrangian is singular the Legendre transformation is not a diffeomorphism, and hence, the resulting Hamiltonian description is not equivalent to the Lagrangian description. To remedy this situation one has to apply the Dirac algorithm for field theories with constraints, \cf \cite{dirac:lqm} or the modern treatment in \cite[Chapter 1.2 and Chapter 24]{thiemann:book}, resulting in two constraints in phase space, the \tit{Hamiltonian} and the \tit{Diffeomorphism constraint} respectively, which are supposed to vanish.

The quantization of these constraints is in general an unsolved problem; however, when the available degrees of freedom are reduced by requiring spherical symmetry for instance, as is the case in cosmology, then the complexity of the constraint equations reduces considerably.

Assuming spherical symmetry for the spatial cross-sections the Diffeomorphism constraint is automatically satisfied and hasn't to be considered any longer. The Hamiltonian constraint can at least formally be quantized leading to a constraint for the possible wave functions, the so-called Wheeler-DeWitt equation, which looks like
\begin{equation}\lae{1.1.7}
H\psi=0,
\end{equation}
where $H$ is a second order hyperbolic operator.

Though special solutions of the Wheeler-DeWitt equation can be found in some circumstances either by trial and error or by an existence proof, these solutions offer no satisfactory answer to the problem of finding a quantum cosmological model.

In quantum theory one always considers a selfadjoint operator in an infinite dimensional Hilbert space which is in general given as the space of complex valued square integrable functions over some measure space.

Thus, a satisfactory treatment of the Wheeler-DeWitt equation requires to solve the equation in a Hilbert space and the solutions have to be associated with a selfadjoint operator.  One might think that the assumption $H$ selfadjoint would suffice, but even if $H$ were selfadjoint and all solutions of \re{1.1.7} could be determined we would face the problem that we had no dynamical development, since the Schr\"odinger equation would make no sense because of \re{1.1.7}.

We shall therefore treat the Wheeler-DeWitt equation as an \tit{implicit eigenvalue equation}, where the cosmological constant plays the role of the eigenvalue either directly or indirectly.

This approach will also reveal that the operator $H$ is not the actual Hamilton operator of the model, instead the corresponding eigenvalue problem will define a symmetric differential operator, and we have to prove that it is selfadjoint.

To our knowledge the Wheeler-DeWitt equation hasn't been considered as an implicit eigenvalue equation before and therefore a Hilbert space approach with a spectral resolution hasn't been achieved either.\footnote{One of the referees has drawn our attention to a paper by Unruh \cite{unruh} in which the Lagrangian is considered in a restricted class of metrics such that the Legendre transformation is a diffeomorphism in the homogeneous case. Thus no constraints have to be imposed on the Hamiltonian and after quantization one could consider the Scr\"odinger equation or the corresponding eigenvalue equation. Unruh proved that the eigenvalues correspond to cosmological constants in classical general relativity.}

DeWitt was the first to quantize the closed Friedmann model with spherical cross-sections in \cite{dewitt:qg}, where he used as a matter Lagrangian not a Lagrangian representing a field, i.e., a Lagrangian which would have to be integrated over the spacetime, but rather a Lagrangian representing finitely many particles resulting in a Hamiltonian where the matter part didn't depend on the scale factor $r$ and hence the Wheeler-DeWitt equation could be written in the form
\begin{equation}
A\psi=E\psi,
\end{equation}
where $A$ is an ordinary differential operator of order two with respect to the variable $r$ and similarly $E$ a Hamiltonian with respect to the finitely many particles $q^i$. Using then a separation of variables, solutions of the Wheeler-DeWitt equation will exist whenever the eigenvalues for the operators $A$ and $E$ coincide. The variable $r$ was supposed to belong to a bounded interval $(0,r_0)$. 

Solutions of a Wheeler-DeWitt equation with a single scalar field are described in \cite[Chapter 8]{kiefer:book} in case of a closed Friedman universe with spherical cross-sections, where the scale factor $r$ belongs  to the unbounded interval $(0,\un)$. Special solutions are constructed with the help of Bessel functions. 

\cvm
We look at  \cq{radially} symmetric spacetimes $N=N^{n+1}$ where the Lorentzian metrics are of the form
\begin{equation}\lae{1.1}
d\bar s^2=-w^2 dt^2+r^2\s_{ij}(x)dx^idx^j;
\end{equation}
here $(\s_{ij})$ is the metric of a spaceform $\so$\footnote{We assume $\so$ to be compact.} with curvature $\tilde\ka$, which could be positive, zero, or negative, and $r$, $w$ are positive functions depending only on $t$, and the Einstein equations are the Euler-Lagrange equation of   the functional \begin{equation}\lae{1.2.6}
J=\int_N(\bar R-2\Lam)+\al_MJ_M,
\end{equation}
where $\bar R$ is the scalar curvature, $\Lam$ a cosmological constant, $\al_M$ a positive coupling constant, and $J_M$ a functional representing matter.

 We shall consider
\begin{equation}
J_M=\int_N\{-\tfrac12\norm{D\f}^2+V(\f)\},
\end{equation}
where $\f$ is a scalar fields map
\begin{equation}
\f:N\ra S
\end{equation}
from $N$ into a compact Riemannian manifold $S=S^m$ with metric $(G_{AB})$, i.e.,
\begin{equation}
J_M=\int_N\{-\tfrac12 \bar g^{\al\bet}\f^A_\al\f^B_\bet G_{AB}+V(\f)\};
\end{equation}
$\f$ is also supposed to be radially symmetric depending only on $t$.

The functional in \re{1.2.6} can be treated as a constrained Hamiltonian problem, \cf \cite{adm:old,dewitt:qg,misner:qc,kiefer:book}, where the Hamiltonian $H$ has to satisfy the Hamiltonian constraint
\begin{equation}
H=0.
\end{equation}

To derive quantum cosmological Friedman models, we therefore  shall quantize the Hamilton function $H$, obtain a Hamilton operator in a suitable Hilbert space $\mc H$ and shall consider only those wave functions $\psi$ satisfying the Wheeler-DeWitt equation
\begin{equation}\lae{1.18}
H\psi=0,
\end{equation}
where we use the same symbol for the Hamilton operator as for the Hamilton function. 

Assuming $V=\const$ the Hamilton operator $H$ is equal to
\begin{equation}\lae{1.11}
\begin{aligned}
H\psi=r^{-1}\tfrac \pa{\pa r}(r\dot \psi)&+r^{-2}(-\tfrac{4}{n^2}a_0\D \psi-\tfrac{(m-1)^2}4\psi)\\
&+\tfrac{16}{n^2}(\bar V+\bar\Lam)r^{2}\psi-\tfrac{16}{n^2}\tilde\ka r^\frac{2(n-2)}{n}\psi,
\end{aligned}
\end{equation} 
where
\begin{equation}\lae{1.12}
a_0=2n(n-1)\al_M^{-1},
\end{equation}
\begin{equation}\lae{1.13}
\bar V=\frac{\al_M}{n(n-1)}V,
\end{equation}
and
\begin{equation}\lae{1.14} 
\bar\Lam= \frac2{n(n-1)}\Lam.
\end{equation}

To find functions in the kernel of $H$, we make  a separation ansatz
\begin{equation}
\psi(r,y^A)=u(r)\h(y).
\end{equation}

Since we assumed $S$ to be compact, $-\D$ has a complete set of eigenfunctions $(\h_i)$ with corresponding eigenvalues $\m_i\ge 0$ such that
\begin{equation}
\lim_i\m_i=\un.
\end{equation}

Let $\h$ be an eigenfunction with eigenvalue $\m$ such that
\begin{equation}\lae{1.25.2}
\bar\m=\tfrac{4}{n^2}a_0\m-\tfrac{(m-1)^2}4\le 0,
\end{equation}
then the resulting differential operator
\begin{equation}
Au=r^{-1}(r\dot u)'+r^{-2}\bar \m u +\tfrac{16}{n^2}(\bar V+\bar\Lam)r^{2}u-\tfrac{16}{n^2}\tilde\ka r^\frac{2(n-2)}{n}u
\end{equation}
is of the form
\begin{equation}
Au=-Bu+\tfrac{16}{n^2}(\bar V+\bar\Lam)r^{2}u-\tfrac{16}{n^2}\tilde\ka r^\frac{2(n-2)}{n}u,
\end{equation}
where $Bu$ is a Bessel operator, i.e., on any finite interval $(0,r_0)$,  $A$ is selfadjoint in a suitable Hilbert space with a complete set of eigenfunctions $(u_i)$ and corresponding eigenvalues $\lam_i$ such that
\begin{equation}
\lim_i\lam_i=-\un.
\end{equation}

To solve the equation 
\begin{equation}\lae{1.29.2}
Au=0
\end{equation}
we distinguish two cases: First, we consider the equation as an implicit eigenvalue problem with respect to the quadratic form
\begin{equation}
K(u)=\tfrac{16}{n^2}\int_I r^3 u^2
\end{equation}
and second, as an implicit eigenvalue problem with respect to the quadratic form
\begin{equation}
K(u)=-\tfrac{16}{n^2}\tilde\ka\int_I r^{\frac{3n-4}n} u^2,
\end{equation}
in which case $\tilde\ka$ has to be negative. In the first case the eigenfunctions have to be defined in a bounded interval $I=(0,r_0)$, and therefore this case is also referred to as the bounded case or the bounded model, while in the second case the eigenfunctions   can be defined in $I=(0,\un)$, i.e., we have an unbounded model.

Let us first consider a bounded interval $I=(0,r_0)$. Writing the equation \re{1.29.2} in the equivalent form
\begin{equation}\lae{1.30}
Bu-\tfrac{16}{n^2}\bar V r^{2}u+\tfrac{16}{n^2}\tilde\ka r^\frac{2(n-2)}{n}u=\tfrac{16}{n^2}\bar\Lam r^{2} u,
\end{equation}
we shall treat it as an eigenvalue problem with eigenvalue $\bar\Lam$. 

Choosing the appropriate Hilbert space we shall show that this problem has countably many eigenvalues $\bar\Lam_i$ and corresponding real eigenfunctions $u_i$ such that
\begin{equation}
\bar\Lam_i<\bar\Lam_{i+1}\qq\A\,i\in\N,
\end{equation}
\begin{equation}
\lim_i\bar\Lam_i=\un,
\end{equation}
and their multiplicities are one.

The right end point $r_0$ of the interval $I$ can be arbitrary, and the eigenvalues $\bar\Lam_i$ as well as the eigenfunctions will depend on its value. To remove the arbitrariness of $r_0$ from the problem, consider a fixed $\bar\Lam$ in the equation \re{1.30}. Then, if either
\begin{equation}\lae{1.30c}
\bar V+\bar\Lam>0\q\wed\q\tilde\ka \;\text{arbitrary},
\end{equation}
or
\begin{equation}\lae{1.31c}
\bar V+\bar\Lam=0\q\wed\q\tilde\ka <0,
\end{equation}
there will be exactly one $r_0>0$ such that $\bar\Lam$ will be the smallest eigenvalue $\bar\Lam_0$
for the eigenvalue problem \re{1.30} in that particular interval.

In case of the unbounded model, let $I=(0,\un)$ and write equation \re{1.29.2} in the form
\begin{equation}\lae{1.35.2}
Bu-\tfrac{16}{n^2}(\bar V+\bar\Lam) r^{2}u=-\tfrac{16}{n^2}\tilde\ka r^\frac{2(n-2)}{n}u,
\end{equation}
where $\bar V+\bar\Lam$ and $\tilde\ka$  are supposed to be negative
\begin{equation}\lae{1.36.2}
\bar V+\bar\Lam<0\q\wed\q \tilde\ka<0.
\end{equation}

Assuming then without loss of generality that $\bar V=0$, we shall show that this eigenvalue problem has countably many solutions $(\bar\Lam_i,u_i)$ such that
\begin{equation}
\bar\Lam_i<\bar\Lam_{i+1}<0,
\end{equation}
\begin{equation}
\lim_{i}\bar\Lam_i=0,
\end{equation}
and their multiplicities are one.

Notice that the present variable $r$ is not identical with the one in \re{1.1}. Writing the metric in \re{1.1} in the form
\begin{equation}\lae{1.30.6}
d\bar s^2=-w^2 dt^2+e^{2f}\s_{ij}dx^idx^j,
\end{equation}
then
\begin{equation}
r=e^{\frac n2 f}.
\end{equation}

However, $r\ra0$ corresponds to $e^{2f}\ra0$, i.e., there will always be a \tit{big bang} singularity. 
\bd
Let $I\su\R[]_+$ be an open interval, not necessarily bounded, such that $0\notin I$, and let $q\in\R[]$. Then we define
\begin{equation}
L^2(I,q)=\set{u\in L^2_{\tup{loc}}(I,\Cc)}{\int_Ir^q\abs u^2<\un}.
\end{equation}
$L^2(I,q)$ is a Hilbert space with scalar product
\begin{equation}
\spd{u_1}{u_2}=\int_I r^qu_1\bar u_2.
\end{equation}
\ed

We are especially interested in $L^2(I,1)$ for $I=(0,r_0)$ and $I=(0,\un)$.

Let $\mc H_0\su L^2(S,\Cc)$ be the finite dimensional subspace spanned by the eigenspaces $E_{\m_i}$ of $-\D=-\D_S$ with eigenvalues $\m_i$ satisfying
\begin{equation}\lae{1.28}
\bar\m_i=\tfrac{4}{n^2}a_0\m_i-\tfrac{(m-1)^2}4\le 0,
\end{equation}
then it is fairly easy to prove:
\bt\lat{1.4}
The Hamilton operator $H$ defined in \re{1.11} is selfadjoint in $L^2((0,r_0),1)\otimes \mc H_0$ for arbitrary values of $\bar V$, $\bar \Lam$ and $\bar \ka$, and also in $L^2(\R[*]_+,1)\otimes\mc H_0$, if in addition
\begin{equation}\lae{1.29}
\bar V+\bar\Lam\le 0\q\wed\q \tilde\ka\ge0,
\end{equation}
or if
\begin{equation}\lae{1.29.3}
\bar V+\bar\Lam< 0\q\wed\q \tilde\ka\q\text{arbitrary}.
\end{equation}
\et

\br
Notice that the dimension of $\mc H_0$ can be fairly large. Indeed, let $G=(G_{AB})$ be a given metric on $S$, and let $\m_i$ be the eigenvalues of the corresponding negative Laplacian, then the eigenvalues of the negative Laplacians corresponding to the metrics 
\begin{equation}
G_\e=(\e^{-2}G_{AB}),\qq\e>0,
\end{equation}
are
\begin{equation}
\m_{\e,i}=\e^2\m_i.
\end{equation}
\er

\bd
(i) Let $I=(0,r_0)$ and $B$ be the Bessel operator
\begin{equation}
Bu=-r^{-1}(r\dot u)'-r^{-2}\bar \m u,\qq \bar\mu\le0.
\end{equation}
 $B$ is defined in the Hilbert space $\mc H_1(\bar\mu)$ which is the completion of $C^\un_c(I)$ with respect to the scalar product
\begin{equation}
\spd uv_1=\int_Ir\dot u \bar{\dot {v}}-\bar\mu\int_I r^{-1}u\bar v,
\end{equation}
such that
\begin{equation}
\spd{Bu}v=\spd uv_1\qq\A\,u,v\in \mc H_1(\bar\mu),
\end{equation}
where the scalar product on the left-hand side is the scalar product in $L^2(I,1)$.

(ii) Let $I=(0,\un)$, then we define the Hilbert space $\mc H_2(\bar\mu)$ as the completion of $C^\un_c(I)$ with respect to the scalar product
\begin{equation}
\spd uv_2=\int_Ir\dot u \bar{\dot {v}}-\bar\mu\int_I r^{-1}u\bar v +\int_I r^3u\bar v.
\end{equation}
\ed

Some of the main results are: 
\bt\lat{1.5}
Let $I=(0,r_0)$ be an arbitrary open interval, let $\mu$ be an eigenvalue of $-\D_S$ such that the eigenspace $E_{\mu}\su\mc H_0$, and let $\h\in E_{\mu}$. Then there are countably many solutions $(\bar\Lam_i,u_i)$ of the eigenvalue problem \re{1.30} with eigenfunctions $u_i\in \mc H_1(\bar\mu)$ such that
\begin{equation}
\bar\Lam_i<\bar\Lam_{i+1}\qq\A\,i\in\N,
\end{equation}
\begin{equation}
\lim_i\bar\Lam_i=\un,
\end{equation}
and their multiplicities are one. The $(u_i)$ form a basis of $\mc H_1(\bar\mu)$ and also of $L^2(I,1)$.
The wave functions
\begin{equation}
\psi_i=u_i\h
\end{equation}
are solutions of
\begin{equation}
H\psi_i=0.
\end{equation}
\et
\bt\lat{1.6}
Let $\Lam, V,\tilde\ka$ be given data such that the conditions \re{1.30c} or \re{1.31c} are satisfied. Then there is exactly one $0<r_0$ such that $\bar\Lam$ is equal to the smallest eigenvalue $\bar\Lam_0$ specified  in the preceding theorem.
\et
\bt\lat{1.7}
Let $I=(0,\un)$ and assume that $\bar V, \bar\Lam$ satisfy the condition in \re{1.36.2} with $\bar V=0$. Let $\mu$ be an eigenvalue of $-\D_S$ such that the eigenspace $E_{\mu}\su\mc H_0$, and let $\h\in E_{\mu}$. Then there are countably many solutions $(\bar\Lam_i,u_i)$ of the eigenvalue problem \re{1.35.2} with eigenfunctions $u_i\in \mc H_2(\bar\mu)$ such that
\begin{equation}
\bar\Lam_i<\bar\Lam_{i+1}<0\qq\A\,i\in\N,
\end{equation}
\begin{equation}
\lim_i\bar\Lam_i=0,
\end{equation}
and their multiplicities are one. 
The wave functions
\begin{equation}
\psi_i=u_i\h
\end{equation}
are solutions of
\begin{equation}
H\psi_i=0.
\end{equation}
The transformed eigenfunctions 
\begin{equation}
\tilde u_i(r)=u_i(\lam_i^{\frac{n}{4(n-1)}}r),
\end{equation}
where
\begin{equation}
\lam_i=(-\bar\Lam_i)^{-\frac {n-1}{n}},
\end{equation}
form a basis of $\mc H_2(\bar\mu)$ and also of $L^2(I,1)$.
\et

\br
The unbounded model is in our opinion the physically most appealing model: first, because there is no arbitrariness about the length of the interval in which the wave functions are defined, second, there is no discrepancy with respect to the maximal radius when it is compared with its classical counterparts,  and third,  for large quantum numbers, and thus for high energies, the absolute values of the corresponding cosmological constants are very small, they tend to zero. This is an invertible relation, i.e., negative cosmological constants close to zero, which correspond to eigenvalues, correspond to high energy levels. 

Given the empirical observations that currently the cosmological constant has a very small absolute value, approximately $10^{-120}$, and also that the universe is accelerating very fast, indicating a large amount of so-called dark energy, then these data could be theoretically explained by the unbounded model.
\er

The paper is organized as follows: \rs{01} contains an overview of our notations and definitions. In \rs{2} the Lagrangian is quantized, while in \rs{3} unitarily equivalent Hamiltonians are derived.

The theorems will be proved   in \rs{4} and \rs{5}.

In \rs{6} we show the existence of a smooth transition from big crunch to big bang under certain circumstances.

In \rs{8} we prove that the implicitly defined eigenvalue problems can be looked at as eigenvalue problems for selfadjoint operators, hence the Schr\"odinger equations for these operators can be solved giving rise to a non-trivial dynamical development of arbitrary superpositions of eigenstates.

The problem of time is addressed in \rs{9}, while in \rs{10}, we compare the quantum models with the corresponding classical Friedman solutions, and we shall see, that only in case of the 
unbounded quantum model there is no discrepancy with respect to the maximal radius when it is compared with its classical counterparts.

\section{Notations and definitions}\las{01}

The main objective of this section is to state the equations of Gau{\ss}, Codazzi,
and Weingarten for spacelike hypersurfaces $M$ in a \di {(n+1)} Lorentzian
manifold
$N$.  Geometric quantities in $N$ will be denoted by
$(\bar g_{ \al \bet}),(\riema  \al \bet \ga \de)$, etc., and those in $M$ by $(g_{ij}), 
(\riem ijkl)$, etc.. Greek indices range from $0$ to $n$ and Latin from $1$ to $n$;
the summation convention is always used. Generic coordinate systems in $N$ resp.
$M$ will be denoted by $(x^ \al)$ \resp $(\x^i)$. Covariant differentiation will
simply be indicated by indices, only in case of possible ambiguity they will be
preceded by a semicolon, i.e., for a function $u$ in $N$, $(u_ \al)$ will be the
gradient and
$(u_{ \al \bet})$ the Hessian, but e.g., the covariant derivative of the curvature
tensor will be abbreviated by $\riema  \al \bet \ga{ \de;\e}$. We also point out that
\begin{equation}
\riema  \al \bet \ga{ \de;i}=\riema  \al \bet \ga{ \de;\e}x_i^\e
\end{equation}
with obvious generalizations to other quantities.

Let $M$ be a \tit{spacelike} hypersurface, i.e., the induced metric is Riemannian,
with a differentiable normal $\n$ which is timelike.

In local coordinates, $(x^ \al)$ and $(\x^i)$, the geometric quantities of the
spacelike hypersurface $M$ are connected through the following equations
\begin{equation}\lae{01.2}
x_{ij}^ \al= h_{ij}\n^ \al
\end{equation}
the so-called \tit{Gau{\ss} formula}. Here, and also in the sequel, a covariant
derivative is always a \tit{full} tensor, i.e.

\begin{equation}
x_{ij}^ \al=x_{,ij}^ \al-\ch ijk x_k^ \al+ \cha  \bet \ga \al x_i^ \bet x_j^ \ga.
\end{equation}
The comma indicates ordinary partial derivatives.

In this implicit definition the \tit{second fundamental form} $(h_{ij})$ is taken
with respect to $\n$.

The second equation is the \tit{Weingarten equation}
\begin{equation}
\n_i^ \al=h_i^k x_k^ \al,
\end{equation}
where we remember that $\n_i^ \al$ is a full tensor.

Finally, we have the \tit{Codazzi equation}
\begin{equation}
h_{ij;k}-h_{ik;j}=\riema \al \bet \ga \de\n^ \al x_i^ \bet x_j^ \ga x_k^ \de
\end{equation}
and the \tit{Gau{\ss} equation}
\begin{equation}
\riem ijkl=- \{h_{ik}h_{jl}-h_{il}h_{jk}\} + \riema  \al \bet\ga \de x_i^ \al x_j^ \bet
x_k^ \ga x_l^ \de.
\end{equation}

Now, let us assume that $N$ is a globally hyperbolic Lorentzian manifold with a
\tit{compact} Cauchy surface. 
$N$ is then a topological product $I\times \mc S_0$, where $I$ is an open interval,
$\mc S_0$ is a compact Riemannian manifold, and there exists a Gaussian coordinate
system
$(x^ \al)$, such that the metric in $N$ has the form 
\begin{equation}\lae{01.7}
d\bar s_N^2=e^{2\psi}\{-{dx^0}^2+\s_{ij}(x^0,x)dx^idx^j\},
\end{equation}
where $\s_{ij}$ is a Riemannian metric, $\psi$ a function on $N$, and $x$ an
abbreviation for the spacelike components $(x^i)$. 
We also assume that
the coordinate system is \tit{future oriented}, i.e., the time coordinate $x^0$
increases on future directed curves. Hence, the \tit{contravariant} timelike
vector $(\x^ \al)=(1,0,\dotsc,0)$ is future directed as is its \tit{covariant} version
$(\x_ \al)=e^{2\psi}(-1,0,\dotsc,0)$.

Let $M=\graph \fv u\so$ be a spacelike hypersurface
\begin{equation}
M=\set{(x^0,x)}{x^0=u(x),\,x\in\mc S_0},
\end{equation}
then the induced metric has the form
\begin{equation}
g_{ij}=e^{2\psi}\{-u_iu_j+\s_{ij}\}
\end{equation}
where $\s_{ij}$ is evaluated at $(u,x)$, and its inverse $(g^{ij})=(g_{ij})^{-1}$ can
be expressed as
\begin{equation}\lae{01.10}
g^{ij}=e^{-2\psi}\{\s^{ij}+\frac{u^i}{v}\frac{u^j}{v}\},
\end{equation}
where $(\s^{ij})=(\s_{ij})^{-1}$ and
\begin{equation}\lae{01.11}
\begin{aligned}
u^i&=\s^{ij}u_j\\
v^2&=1-\s^{ij}u_iu_j\equiv 1-\abs{Du}^2.
\end{aligned}
\end{equation}
Hence, $\graph u$ is spacelike if and only if $\abs{Du}<1$.

The covariant form of a normal vector of a graph looks like
\begin{equation}
(\n_ \al)=\pm v^{-1}e^{\psi}(1, -u_i).
\end{equation}
and the contravariant version is
\begin{equation}
(\n^ \al)=\mp v^{-1}e^{-\psi}(1, u^i).
\end{equation}
Thus, we have
\br Let $M$ be spacelike graph in a future oriented coordinate system. Then the
contravariant future directed normal vector has the form
\begin{equation}
(\n^ \al)=v^{-1}e^{-\psi}(1, u^i)
\end{equation}
and the past directed
\begin{equation}\lae{01.15}
(\n^ \al)=-v^{-1}e^{-\psi}(1, u^i).
\end{equation}
\er

In the Gau{\ss} formula \re{01.2} we are free to choose the future or past directed
normal, but we stipulate that we always use the past directed normal for reasons
that we have explained in \ci[Section 2]{cg:indiana}.

Look at the component $ \al=0$ in \re{01.2} and obtain in view of \re{01.15}

\begin{equation}\lae{01.16}
e^{-\psi}v^{-1}h_{ij}=-u_{ij}- \cha 000\mspace{1mu}u_iu_j- \cha 0j0
\mspace{1mu}u_i- \cha 0i0\mspace{1mu}u_j- \cha ij0.
\end{equation}
Here, the covariant derivatives are taken with respect to the induced metric of
$M$, and
\begin{equation}
-\cha ij0=e^{-\psi}\bar h_{ij},
\end{equation}
where $(\bar h_{ij})$ is the second fundamental form of the hypersurfaces
$\{x^0=\const\}$.

An easy calculation shows
\begin{equation}
\bar h_{ij}e^{-\psi}=-\tfrac{1}{2}\dot\s_{ij} -\dot\psi\s_{ij},
\end{equation}
where the dot indicates differentiation with respect to $x^0$.

\section{The quantization of the Lagrangian}\las{2}

Consider the functional
\begin{equation}\lae{1.10}
J=\int_N(\bar R-2\Lam)+\al_M\int_N\{-\tfrac12 \bar g^{\al\bet}\f^A_\al\f^B_\bet G_{AB}+V(\f)\},
\end{equation}
where the $(n+1)$-dimensional spacetimes $N$ have a metric of the form \fre{1.30.6}. The time coordinate $t$ is supposed to belong to a fixed interval $I=(a,b)$, bounded or unbounded, the actual size of which is unimportant, since we are only interested in the first variation of the functional with respect to compact variations. 

To express the scalar curvature $\bar R$ in terms of $f$ we use the contracted Gau{\ss} equation. Let $M_t$ be the spacelike hypersurface
\begin{equation}
M_t=\set{p\in N}{x^0(p)=t},
\end{equation}
i.e., $M_t$ is a coordinate slice and $x^0$ is the future directed time function, which we also called $t$, but here $t\in (a,b)$ is an arbitrary but fixed value.

The induced metric of $M_t$ is
\begin{equation}\lae{2.2}
g_{ij}=e^{2f}\s_{ij}.
\end{equation}

Because of the radial symmetry, the principal curvatures of the hypersurfaces are all identical and can be expressed as
\begin{equation}
\ka_i\equiv\ka=-f'w^{-1}.
\end{equation}

Let $H$ be the mean curvature
\begin{equation}
H=\sum_i\ka_i=n\ka
\end{equation}
and $\abs A^2$ be defined by
\begin{equation}
\abs A^2=\sum_i\ka_i^2=n\ka^2,
\end{equation}
and let $R$ be the scalar curvature of the $M_t$, then 
\begin{equation}\lae{2.6}
R=-\{H^2-\abs A^2\}+\bar R+2\bar R_{\al\bet}\nu^\al\nu^\bet,
\end{equation}
\cf \cite[equation (1.1.41)]{cg:cp}.

From \re{2.2} we deduce that $g_{ij}$ is a constant multiple of the metric $\s_{ij}$ of the spaceform $\so$, hence
\begin{equation}
R=e^{-2f}R_{\so}=n(n-1)\tilde\ka e^{-2f}.
\end{equation}

It remains to express $\bar R_{\al\bet}\nu^\al\nu^\bet$.

The easiest way to achieve this is by writing the metric in \fre{1.30.6} in its conformal time form
\begin{equation}\lae{3.9.7} 
d\bar s^2=e^{2f}\{-d\tau^2+\s_{ij}dx^idx^j\}
\end{equation}
such that
\begin{equation}\lae{2.10.6}
\df \tau{t}=we^{-f}.
\end{equation}

The ambient metric $\bar g_{\al\bet}$ is now conformal to a product metric $g_{\al\bet}$
\begin{equation}
\bar g_{\al\bet}=e^{2f} g_{\al\bet}.
\end{equation}
The Ricci tensors $\bar R_{\al\bet}$ and $R_{\al\bet}$ are then related by the formula
\begin{equation}
\bar R_{\al\bet}=R_{\al\bet}-(n-1)[f_{\al\bet}-f_\al f_\bet]-g_{\al\bet}[\D f+(n-1)\norm{Df}^2],
\end{equation}
where the covariant derivatives of $f$ are taken with respect to the metric $g_{\al\bet}$.

Since $R_{00}=0$, we immediately conclude
\begin{equation}
\bar R_{00}=-nf'',
\end{equation}
and hence
\begin{equation}
\bar R_{\al\bet}\nu^\al\nu^\bet=-nf''e^{-2f},
\end{equation}
for the past directed normal $\nu$ is equal to
\begin{equation}
\nu=(\nu^\al)=-e^{-f}(1,0,\ldots,0).
\end{equation}

Switching back to $t$ as time coordinate, we deduce from \re{2.10.6} and \re{2.6}
\begin{equation}
\begin{aligned}
\bar R&=e^{-2f}R_\so+H^2-\abs A^2-2\bar R_{\al\bet}\nu^\al\nu^\bet\\
&=n(n-1)\tilde\ka e^{-2f}+n(n-1)\abs{f'}^2w^{-2}+2n\df {}t\{w^{-1}e^ff'\}w^{-1}e^{-f}.
\end{aligned}
\end{equation}

The volume element of $N$ is
\begin{equation}
\sqrt{\abs{\bar g}}=we^{nf}\sqrt\s,
\end{equation}
where $\s=\det(\s_{ij})$. Since all terms in the functional do not depend on $(x^i)$, integration over $\so$ is simply a multiplication by the volume of $\so$ and without loss of generality we shall set volume of $\so$ to be equal to one
\begin{equation}
\abs\so=1.
\end{equation}

The functional can therefore be written as
\begin{equation}
\begin{aligned}
\int_a^b\{&n(n-1)\tilde\ka e^{(n-2)f}w+n(n-1)\abs{f'}^2e^{nf}w^{-1}\\
&+2n\df {}t\{w^{-1}e^ff'\}e^{(n-1)f}-2\Lam e^{nf}w\}\\
&+\al_M\int_a^b\{\tfrac12\norm{\dot\f}^2e^{nf}w^{-1}-V(\f)e^{nf}w\}
\end{aligned}
\end{equation}

To eliminate the term involving the second derivatives of $f$, we observe that
\begin{equation}
\frac d{dt}(w^{-1}f'e^{nf})=\df {}t\{w^{-1}e^ff'\}e^{(n-1)f}+(n-1)\abs{f'}^2e^{nf}w^{-1}.
\end{equation}
The expression on the left-hand side is a total derivative, i.e., its first variation with respect to compact variations vanishes, hence we obtain an equivalent functional, still denoted by $J$,
\begin{equation}\lae{2.18}
\begin{aligned}
\int_a^b\{&n(n-1)\tilde\ka e^{(n-2)f}w-n(n-1)\abs{f'}^2e^{nf}w^{-1}
-2\Lam e^{nf}w\}\\
&+\al_M\int_a^b\{\tfrac12\norm{\dot\f}^2e^{nf}w^{-1}-V(\f)e^{nf}w\}.
\end{aligned}
\end{equation}

Before we apply the Legendre transformation let us normalize the functional by dividing the whole expression by $n(n-1)$. Denoting the resulting functional still by $J$, we obtain
\begin{equation}
\begin{aligned}
\int_a^b\{&\tilde\ka e^{(n-2)f}w-\abs{f'}^2e^{nf}w^{-2}
-\tfrac2{n(n-1)}\Lam e^{nf}\}w\\
&+\int_a^b\{\frac{\al_M}{2n(n-1)}\norm{\dot\f}^2e^{nf}w^{-1}-\frac{\al_M}{n(n-1)}V(\f)e^{nf}w\}.
\end{aligned}
\end{equation}

Define $\bar V$, $\bar\Lam$ as in \re{1.13},\fre{1.14}, set
\begin{equation}
\bar G_{AB}=\frac{\al_M}{2n(n-1)} G_{AB},
\end{equation}
\begin{equation}
(\widetilde G_{ab})=\begin{pmatrix}
-1&0\\[\cma]
0&\bar G_{AB}
\end{pmatrix}e^{nf}
\end{equation}
and 
\begin{equation}
(y^a)=(y^0,y^A)=(f,\f^A),
\end{equation}
then $J$ can be expressed as
\begin{equation}
\begin{aligned}
J=\int_a^bL=\int_a^bw\{\widetilde G_{ab}\dot y^a\dot y^bw^{-2}-(\bar V+\bar \Lam)e^{nf}+\tilde\ka e^{(n-2)f}\}.
\end{aligned}
\end{equation}

Applying now the Legendre transformation we obtain the Hamiltonian $\tilde H$
\begin{equation}
\begin{aligned}
\tilde H=\tilde H(w,y^a,p_a)&=\dot y^a\pde L{\dot y^a}-L\\
&=\{\wt G_{ab} \dot y^a\dot y^bw^{-2} +(\bar V+\bar\Lam)e^{nf}-\tilde\ka e^{(n-2)f}\}w\\
&=\{\tfrac14\wt G^{ab}p_ap_b+(\bar V+\bar\Lam)e^{nf}-\tilde\ka e^{(n-2)f}\}w\\
&\equiv Hw,
\end{aligned}
\end{equation}
and the Hamiltonian constraint requires  
\begin{equation}
H(y^a,p_a)=0.
\end{equation}

Canonical quantization stipulates to replace the momenta $p_a$ by
\begin{equation}
p_a=\frac{\hbar}i\frac{\pa}{\pa y^a}.
\end{equation}

Hence, using the convention $\hbar=1$, we conclude that the Hamilton operator $H$ is equal to
\begin{equation}
H=-\tfrac14\tilde\D+(\bar V+\bar\Lam)e^{nf}-\tilde\ka e^{(n-2)f}.
\end{equation}

Note that the metric $\wt G_{ab}$ is a Lorentz metric, i.e., $H$ is hyperbolic.

Let $\psi=\psi(y)$ be a smooth function then
\begin{equation}
\tilde\D\psi=\frac1{\sqrt\abs{\wt G}}\frac\pa{\pa y^a}\bigg(\sqrt{\abs{\wt G}}\wt G^{ab}\psi_b\bigg).
\end{equation}

Now
\begin{equation}
\abs{\wt G}=e^{(m+1)nf}\abs {\bar G}
\end{equation}
and hence
\begin{equation}\lae{2.33.6}
-\tilde\D\psi=e^{-\frac{(m+1)n}2f}\frac\pa{\pa y^0}\big(e^{\frac{(m-1)n}2f}\pde\psi{y^0}\big)-e^{-nf}\bar\D\psi,
\end{equation}
where $\bar\D$ is the Laplacian with respect to $\bar G_{AB}$.

We now define a new variable
\begin{equation}\lae{2.30}
r=e^{\frac{n}2f}
\end{equation}
and conclude
\begin{equation}
-\tilde\D\psi =\tfrac{n^2}4r^{-m}\pde{}r(r^m\pde\psi{r})-r^{-2}\bar\D\psi.
\end{equation}

Thus we have proved:
\bt
If $m\ge 1$, i.e., if a matter Lagrangian is involved, the Hamilton operator has the form
\begin{equation}\lae{2.31}
\begin{aligned}
H\psi=\tfrac{n^2}{16}r^{-m}\pde{}r(r^m\pde\psi{r})-r^{-2}\tfrac14\bar\D\psi+(\bar V+\bar\Lam)r^{2}\psi-\tilde\ka r^\frac{2(n-2)}{n}\psi.
\end{aligned}
\end{equation}
If no matter Lagrangian has been considered, which is tantamount to $m=0$, $H$ is equal to
\begin{equation}\lae{2.32}
H\psi=\tfrac{n^2}{16}\Ddot\psi+\bar\Lam r^{2}\psi-\tilde\ka r^\frac{2(n-2)}{n}\psi.
\end{equation}
\et
\br
In both cases $r=0$ represents a singularity, since the spacetime metric becomes singular in $r=0$.
\er

\section{Equivalent Hamiltonians}\las{3}

First, we shall divide the Hamiltonian in \re{2.31} by $\tfrac{n^2}{16}$ without changing its symbol such that
\begin{equation}\lae{3.1}
\begin{aligned}
H\psi&=r^{-m}\pde{}r(r^m\pde\psi{r})-r^{-2}\tfrac{4}{n^2}\bar\D\psi\\
&\q+\tfrac{16}{n^2}(\bar V+\bar\Lam)r^{2}\psi-\tfrac{16}{n^2}\tilde\ka r^\frac{2(n-2)}{n}\psi.
\end{aligned}
\end{equation}

Replace $-\tfrac{4}{n^2}\bar\D\psi$ by a non-negative constant $\mu$ and consider the equation
\begin{equation}\lae{3.2.6}
r^{-m}\pde{}r(r^m\pde v{r})+r^{-2}\mu v +\lam v=0
\end{equation}
or its equivalent form
\begin{equation}\lae{3.3}
\pde{}r(r^m\pde v{r})+r^{m-2}\mu v+r^m\lam v=0,
\end{equation}
where $v=v(r)$ and $\lam=\lam(r)$.

If $m\ge 1$, define a new possible solution $u$ by 
\begin{equation}
v=r^{-\frac{m-1}2}u,
\end{equation}
then the left-hand side of \re{3.3} is transformed to
\begin{equation}
r^{\frac{m+1}2}\{\Ddot u +r^{-1}\dot u+ r^{-2}[\mu-\tfrac{(m-1)^2}4]u+\lam u\}.
\end{equation}

The operator inside the braces can be written as
\begin{equation}\lae{3.6}
r^{-1}\pde{}r(r\dot u)+r^{-2}\bar\mu u +\lam u,
\end{equation}
where
\begin{equation}
\bar\mu=\mu -\tfrac{(m-1)^2}4.
\end{equation}

Let $A$ be the operator in \re{3.2.6} and $\hat A$ be the operator in \re{3.6}, then there holds:  
\bl
Let $I\su \R[*]_+$ be an open interval  and let
\begin{equation}
\f:L^2(I,m)\ra L^2(I,1)
\end{equation}
be the linear map
\begin{equation}
\f(v)=u=r^{\frac{m-1}2}v.
\end{equation}
Then $\f$ is unitary and, if $A$ \resp $\hat A$ are defined in $C^\un_c(I)$, there holds
\begin{equation}\lae{3.10}
A=\f^{-1}\circ \hat A\circ \f.
\end{equation}
\el
\bp
(i) Let $v_i\in L^2(I,m)$, $i=1,2$, then
\begin{equation}
\int_Ir^mv_1\bar v_2=\int_Iru_1\bar u_2,
\end{equation}
hence $\f$ is unitary.

\cvm
(ii) To prove \re{3.10}, let $v_i\in C_c^\un(I)$, $i=1,2$. We only need to test the main part of $A$, i.e.,
\begin{equation}\lae{3.12}
\begin{aligned}
\spd{r^{-m}\pde{}r(r^m\dot v_1)}{v_2}&=\int_I\pde{}r(r^m\dot v_1)\bar v_2= \int_I\pde{}r(r^m\dot v_1)r^{-\frac{m-1}2}\bar u_2\\[\cma]
&=\int_I\{-r^{\frac{m+1}2}\dot v_1 \bar {\dot u}_2+\frac{m-1}2r^{\frac {m-1}2}\dot v_1\dot\bar u_2\}.
\end{aligned}
\end{equation}

Now we have
\begin{equation}
\dot v_1=r^{-\frac{m-1}2}\dot u_1-\tfrac{m-1}2 r^{-\frac{m+1}2}u_1
\end{equation}
and thus, the right-hand side of \re{3.12} is equal to
\begin{equation}
\begin{aligned}
\int_I&\{-r\dot u_1\bar{\dot u}_2+\tfrac{m-1}2(u_1\bar{\dot u}_2+\dot u_1\bar u_2)-\tfrac{(m-1)^2}4 r^{-1}u_1\bar u_2\}\\[\cma]
&=\int_I\{-r\dot u_1\bar{\dot u}_2-\tfrac{(m-1)^2}4r^{-1}u_1\bar u_2\},
\end{aligned}
\end{equation}
hence the result, and we may view the Hamiltonian in \fre{1.11} as unitarily equivalent to the original Hamiltonian in \fre{2.31}.
\ep
\section{Proof of \rt{1.4}}\las{4}

Let $\mc H_0\su L^2(S)$ be the finite dimensional subspace spanned by the eigenfunctions of $-\D_S$ such that the corresponding eigenvalues satisfy \fre{1.28}. Then we conclude
\begin{equation}
\tfrac{4}{n^2}a_0\int_S\abs{D\h}^2-\tfrac{(m-1)^2}4\int_S\abs{\h}^2\le 0\qq\A\,\h\in\mc H_0,
\end{equation}
and we deduce further that the quadratic form
\begin{equation}\lae{4.2}
\spd{H\psi}\psi\le c\norm\psi^2\qq\A\,\psi\in C^\un_c(I)\otimes\mc H_0,
\end{equation}
if either $I=(0,r_0)$ and $\bar V,\bar\Lam$ and $\tilde\ka$ are arbitrary, or if $I=\R[*]_+$ and the sign conditions in \fre{1.29} are satisfied.

If $I=(0,\un)$ and if $\bar V,\bar\Lam$ satisfy the condition in \fre{1.36.2}, then, for any $\e>0$ there exists a constant $c_\e$ such that
\begin{equation}
\int_0^\un r^{\frac{3n-4}n}\abs u^2\le \e \norm u^2_2+c_\e\norm u^2\qq\A\, u\in H_2(\bar\mu),
\end{equation}
 where $\norm{\cdot}$ is the norm in $L^2(I,1)$, hence inequality \re{4.2} is also valid in this case.
 
 The preceding inequality follows from  a compactness theorem of Lions and Magenes, see the proof of \frl{5.4} for details, which can be applied because of \frl{5.7}.

$H$ is therefore semibounded and its Friedrichs extension is selfadjoint, hence $H$ is essentially selfadjoint; notice that the Friedrichs extension of a symmetric differential operator can be naturally identified with this operator \cf \frd{7.3} and  the proof of \frt{7.4} for details in a similar situation.

\section{Proof of \rt{1.5}, \rt{1.6} and \rt{1.7}}\las{5}
\subsection{Proof of \rt{1.5}}\lass{5.1}
Let $I=(0,r_0)$ be a fixed interval and let us consider the equation \fre{1.30} for real valued functions $u$.
It is well-known that the Bessel operator 
\begin{equation}
Bu=-r^{-1}(r\dot u)'-r^{-2}\bar \m u,\qq \bar\mu\le0,
\end{equation}
is a selfadjoint, positive definite operator in $L^2(I,1)$---we consider only real valued functions. $B$ is defined in the Hilbert space $\mc H_1(\bar\mu)$ which is the completion of $C^\un_c(I)$ with respect to the scalar product
\begin{equation}
\spd uv_1=\int_Ir\dot u\dot v-\bar\mu\int_I r^{-1}uv,
\end{equation}
such that
\begin{equation}
\spd{Bu}v=\spd uv_1\qq\A\,u,v\in \mc H_1(\bar\mu),
\end{equation}
where the scalar product on the left-hand side is the scalar product in $L^2(I,1)$.

\bl
Functions $u\in \mc H_1(\bar\mu)$ have boundary values zero in $r_0$, i.e., $u(r_0)=0$, while $u(0)=0$ is in general only valid, if $\bar\mu<0$. In case $\bar\mu=0$, there holds
\begin{equation}
\lim_{r\ra0} r^\e u(r)=0
\end{equation}
for any $\e>0$.
\el

\bp
We first point out that any $u\in\mc H_1(\bar\mu)$ is continuous in $I$, even continuous in $(0,r_0]$, and any convergence in the Hilbert space norm implies convergence in $C^0([\de,r_0])$, for arbitrary $\de>0$, hence we conclude $u(r_0)=0$.

\cq{$\bar\mu<0$}\q Then any $u\in\mc H_1(\bar\mu)$ satisfies
\begin{equation}
\int_Ir^{-1}\abs u^2<\un.
\end{equation}
Let $u\in C^\un_c(I)$ and $0<\de<r_0$, then
\begin{equation}
\tfrac12 u^2(\de)\le \int_0^{\de}\abs u\abs{\dot u}\le\bigg(\int_0^{\de}r^{-1}u^2\bigg)^{\frac12} \bigg(\int_0^{\de}r\dot u^2\bigg)^{\frac12},
\end{equation}
and this inequality will also be valid for arbitrary $u\in \mc H_1(\bar\mu)$, hence the result.

\cvm
\cq{$\bar\mu=0$}\q Let $u\in C^\un_c(I)$, $\e>0$, and $0<\de<r_0$, then
\begin{equation}
u(\de)=-\int_\de^{r_0}\dot u,
\end{equation}
and thus
\begin{equation}\lae{5.8}
\de^\e \abs{u(\de)}\le \de^\e(\log r_0-\log\de)^{\tfrac12}\bigg(\int_0^{r_0}r\dot u^2\bigg)^{\tfrac12}.
\end{equation}
This inequality will also be valid for any $u\in\mc H_1(\bar\mu)$, hence the result. 
\ep

\bl\lal{5.3}
Let $K$ be the quadratic form
\begin{equation}\lae{5.9.6}
K(u)=\tfrac{16}{n^2}\int_I r^3 u^2,
\end{equation}
then $K$ is compact in $\mc H_1(\bar\mu)$, i.e., 
\begin{equation}
u_i\whc{\mc H_1(\bar\mu)} u\q\im\q K(u_i)\ra K(u),
\end{equation}
and positive definite, i.e.,
\begin{equation}
K(u)>0\qq\A\,u\ne 0.
\end{equation}
\el
\bp
We may assume that the weak limit $u=0$. Let $0<\de<r_0$, then
\begin{equation}
\lim \tfrac{16}{n^2}\int_\de^{r_0} r^3 u_i^2=0
\end{equation}
and 
\begin{equation}
\limsup \tfrac{16}{n^2}\int_0^{\de} r^3 u_i^2\le c\de,
\end{equation}
in view of \re{5.8}, hence the compactness result.

The positive definiteness is obvious.
\ep

\bl\lal{5.4}
There exist positive constants $c_0$ and $c_1$ such that 
\begin{equation}
\begin{aligned}
c_1\norm u^2_1\le \spd{Bu}u +\spd{-\tfrac{16}{n^2}\bar V r^{2}u+\tfrac{16}{n^2}\tilde\ka r^\frac{2(n-2)}{n}u}u +c_0K(u)
\end{aligned}
\end{equation}
for all $u\in \mc H_1(\bar\mu)$, where the norm on the left-hand side is the norm in $\mc H_1(\bar\mu)$.
\el
\bp
This follows immediately from \re{5.8}, \rl{5.3} and a well-known compactness theorem of J.L. Lions and E. Magenes, \cf \cite[Theorem 16.4]{lions:book}, which says in the present situation that for any $\e>0$ there exists $c_\e$ such that for any $u\in\mc H_1(\bar\mu)$
\begin{equation}
\begin{aligned}
\abs{\spd{-\tfrac{16}{n^2}\bar V r^{2}u+\tfrac{16}{n^2}\tilde\ka r^\frac{2(n-2)}{n}u}u}\le \e\spd{Bu}u+c_\e K(u).
\end{aligned}
\end{equation}
\ep

The eigenvalue problem
\begin{equation}
Bu-\tfrac{16}{n^2}\bar V r^{2}u+\tfrac{16}{n^2}\tilde\ka r^\frac{2(n-2)}{n}u=\bar\Lam \tfrac{16}{n^2}r^{2}u,
\end{equation}
or equivalently,
\begin{equation}
\spd{Bu-\tfrac{16}{n^2}\bar V r^{2}u+\tfrac{16}{n^2}\tilde\ka r^\frac{2(n-2)}{n}u}v=\bar\Lam K(u,v)\q\A\, v\in\mc H_1(\bar\mu),
\end{equation}
where $K(u,v)$ is the bilinear form associated with $K$, then has countably many solutions $(\bar\Lam_i,u_i)$, $u_i\in\mc H_1(\bar\mu)$, with the properties
\begin{equation}\lae{5.18}
\bar\Lam_i<\bar\Lam_{i+1}\qq\A\,i\in\N,
\end{equation}
\begin{equation}
\lim_i\bar\Lam_i=\un,
\end{equation}
\begin{equation}
K(u_i,u_j)=\de_{ij},
\end{equation}
the $(u_i)$ are a Hilbert space basis in $\mc H_1(\bar\mu)$, and the eigenspaces are one dimensional.

This result follows from a general existence result for eigenvalue problems of this kind which goes back to Courant-Hilbert, \cf \cite{cg:eigenwert}, while the strict inequality in \re{5.18} and the property that the multiplicity of the eigenvalues is one is due to the fact that we are dealing with a linear second order ODE, hence the kernel is two dimensional and the boundary condition $u(r_0)=0$ defines a one dimensional subspace.

\subsection{Proof of \rt{1.6}}
Let $Q$ be the quadratic form
\begin{equation}
Q(u)=\spd{Bu-\tfrac{16}{n^2}(\bar V+\bar\Lam) r^{2}u+\tfrac{16}{n^2}\tilde\ka r^\frac{2(n-2)}{n}u}u.
\end{equation}
It suffices to prove that
\begin{equation}\lae{5.22}
\inf\set{Q(u)}{u\in C^\un_c(0,r_0), K(u)=1}>0
\end{equation}
for small $r_0$, and
\begin{equation}\lae{5.23}
\inf\set{Q(u)}{u\in C^\un_c(0,r_0), K(u)=1}<0
\end{equation}
for large $r_0$, where $K$ is the quadratic form in \re{5.9.6}, since the smallest eigenvalue $\bar\Lam_0(r_0)$ depends continuously on $r_0$, as one easily checks.

The claim \re{5.22} follows immediately from the estimate in \re{5.8} by choosing $\de=r_0$.

To prove \re{5.23}, let $r_0$ be large and let $\h\in C^\un_c(\tfrac12,3)$ be a cut-off function such that
\begin{equation}
\h(r)=1\qq\A\, r\in (1,2),
\end{equation}
and define
\begin{equation}
u=\h(\tfrac{3r}{r_0}).
\end{equation}
Then $u\in C^\un_c(0,r_0)$ and
\begin{equation}
\int_0^{r_0}r\dot u^2\le c r_0
\end{equation}
while
\begin{equation}
\int_0^{r_0}r^{\frac{3n-4}n}u^2\sim cr_0^{\frac{4(n-1)}{n}}
\end{equation}
and
\begin{equation}
\int_0^{r_0}r^3u^2\sim c r_0^4,
\end{equation}
which completes the proof of \rt{1.6}.

\br
When we set $V=0$ then \rt{1.6} implies  that for any positive cosmological constant $\Lam$ there is a unique $r_0>0$ such that $\bar\Lam$ is the smallest eigenvalue of the corresponding eigenvalues in $(0,r_0)$, where the sign of $\tilde\ka$ can be arbitrary. In case $\tilde\ka<0$, even $\Lam=0$ can be looked at as the smallest eigenvalue by choosing $r_0$ appropriately.
\er

\subsection{Proof of \rt{1.7}} 
Let $I=(0,\un)$ and let us consider the equation \fre{1.35.2}, where $\bar V, \bar\Lam$ and $\tilde\ka$ satisfy the conditions \re{1.36.2} with $\bar V=0$. The arguments are similar as in the proof of \rt{1.5} in \frss{5.1}.  
\bl\lal{5.7}
Let $K$ be the quadratic form
\begin{equation}
K(u)=-\tilde\ka \tfrac{16}{n^2}\int_I r^{\frac{3n-4}n} u^2,
\end{equation}
then $K$ is compact in $\mc H_2(\bar\mu)$, i.e., 
\begin{equation}
u_i\whc{\mc H_2(\bar\mu)} u\q\im\q K(u_i)\ra K(u),
\end{equation}
and positive definite, i.e.,
\begin{equation}
K(u)>0\qq\A\,u\ne 0.
\end{equation}
\el
\bp
We may assume that the weak limit $u=0$. Let $0<\rho<\un$, then
\begin{equation}
\lim \tfrac{16}{n^2}\int_0^{\rho} r^{\frac{3n-4}n} u_i^2=0
\end{equation}
and 
\begin{equation}
\limsup \tfrac{16}{n^2}\int_\rho^{\un} r^{\frac{3n-4}n} u_i^2\le c\rho^{-\frac4{n}}\int_\rho^\un r^3u_i^2\le c\rho^{-\frac4{n}},
\end{equation}
in view of the definition of the norm in $\mc H_2(\bar\mu)$, hence the compactness result.

The positive definiteness is obvious.
\ep
The analogue of \frl{5.4} is trivially satisfied, in view of the assumption \fre{1.36.2}:
\bl\lal{5.8.2}
There exists a positive constant  $c_1$ such that
\begin{equation}
\begin{aligned}
c_1\norm u^2_2\le \spd{Bu}u +\spd{\tfrac{16}{n^2} r^{2}u}u 
\end{aligned}
\end{equation}
for all $u\in \mc H_2(\bar\mu)$, where the norm on the left-hand side is the norm in $\mc H_2(\bar\mu)$.
\el

Arguing as at the end of \frss{5.1} we then conclude:
The eigenvalue problem
\begin{equation}\lae{5.35.5}
Bu+\tfrac{16}{n^2}r^{2}u=-\lam\tfrac{16}{n^2}\tilde\ka r^\frac{2(n-2)}{n}u,
\end{equation}
or equivalently,
\begin{equation}
\spd{Bu+\tfrac{16}{n^2} r^{2}u}v=\lam  K(u,v)\q\A\, v\in\mc H_2(\bar\mu),
\end{equation}
where $K(u,v)$ is the bilinear form associated with $K$, has countably many solutions $(\lam_i,\tilde u_i)$, $\tilde u_i\in\mc H_2(\bar\mu)$, with the properties
\begin{equation}
\lam_i<\lam_{i+1}\qq\A\,i\in\N,
\end{equation}
\begin{equation}
\lim_i\lam_i=\un,
\end{equation}
\begin{equation}
K(\tilde u_i,\tilde u_j)=\de_{ij},
\end{equation}
the $(\tilde u_i)$ are a Hilbert space basis in $\mc H_2(\bar\mu)$, and the eigenspaces are one dimensional.

The functions
\begin{equation}\lae{6.40.7}
u_i(r)=\tilde u_i(\lam_i^{-\frac n{4(n-1)}} r)
\end{equation}
then satisfy the equation 
\begin{equation}
Bu_i+\tfrac{16}{n^2}\lam_i^{-\frac{n}{n-1}}r^{2}u_i=-\tfrac{16}{n^2}\tilde\ka r^\frac{2(n-2)}{n} u_i
\end{equation}
and they are mutually orthogonal with respect to the bilinear form
\begin{equation}
\int_Ir^3 uv,
\end{equation}
as one easily checks.

Furthermore, there holds:
\bl
Let $(\lam,u) \in \R[*]_+\times \mc H_2(\bar\mu)$,  be a solution of
\begin{equation}
Bu+\tfrac{16}{n^2}\lam^{-\frac{n}{n-1}}r^{2}u=-\tfrac{16}{n^2}\tilde\ka r^\frac{2(n-2)}{n} u,
\end{equation}
then there exists $i$ such that 
\begin{equation}
\lam=\lam_i\q\wed\q u\in\langle u_i\rangle.
\end{equation}
\el
\bp
Define
\begin{equation}
\tilde u(r)=u(\lam ^{\frac n{4(n-1)}}r),
\end{equation}
then the pair $(\lam,\tilde u)$ is a solution of the equation \re{5.35.5}, hence the result.
\ep

\section{Transition from big crunch to big bang}\las{6}
In the previous sections we supposed the singularity $r=0$ to lie in the past, i.e., assuming a future oriented coordinate system in \fre{1.30.6}, $f'$ should be positive.

Of course we could just as well assumed $f'<0$, then $r=0$ would be a future singularity and the relation $f=\log r$ would have been replaced by 
\begin{equation}
f=\log(-r),
\end{equation}
where $r$ is negative.

A similar consideration could have been used in \fre{2.30}, which had to be replaced by
\begin{equation}
-r=e^{\frac{n}2f},\q r<0,
\end{equation}
resulting in a Hamilton operator as in \re{2.31} and \fre{2.32} with the exception that the wave functions would be defined in $(-r_0,0)$, $r_0>0$, or in $(-\un,0)$,  and that in the coefficients on the right-hand side of \re{2.31} \resp \re{2.32} $r$ should be replaced by $-r$, where $r<0$. The singularity $r=0$ would then be a big crunch singularity. Notice that after quantization we use the sign of $r$ to determine  and to define, if the singularity is a big bang, $r>0$, or a big crunch, $r<0$, see \rs{9} for more details.

In order to define and prove the existence of a smooth transition from big crunch to big bang, we consider the variable transformation  
\begin{equation}
r=e^f
\end{equation}
in \fre{2.33.6}, then the equation \re{2.33.6} will be transformed to
\begin{equation}
-\tilde\D\psi=r^{-(n-2)}\{r^{-\frac{(m-1)n+2}2}\pde {}r(r^{\frac{(m-1)n+2}2}\dot\psi)\}-r^{-n}\bar\D\psi.
\end{equation}

Hence, the equation
\begin{equation}
H\psi=0
\end{equation}
is equivalent to
\begin{equation}
\hat H\psi\equiv r^{-p}\pde {}r(r^p\dot\psi)-r^{-2}\bar\D\psi+4(\bar V+\bar\Lam)r^{2(n-1)}\psi-4\tilde\ka r^{2(n-2)}\psi=0,
\end{equation}
where
\begin{equation}
p=\tfrac{(m-1)n+2}2.
\end{equation}

After a separation of variables
\begin{equation}
\psi=u\h,
\end{equation}
let us introduce another equivalent Hamiltonian: Let $v$ a solution of \fre{3.3}, but define $u$ by
\begin{equation}\lae{6.3}
v=r^{-\frac p2}u,\q  r>0,
\end{equation}
or by
\begin{equation}
v=(-r)^{-\frac p2}u,\q r<0,
\end{equation}
in case $r$ is negative.

Then the left-hand side of \fre{3.3} is transformed to
\begin{equation}
\abs r^{\frac p2}\{\Ddot u+r^{-2}[\mu-\tfrac{p(p-2)}4]u+\lam u\},
\end{equation}
i.e., the unitarily equivalent Hamilton operator would be of the form
\begin{equation}\lae{6.6}
\begin{aligned}
\Ddot u &+r^{-2}[a_0\mu -\tfrac{p(p-2)}4]u 
+4(\bar V+\bar\Lam)r^{2(n-1)}u-4\tilde \ka r^{2(n-2)}u, 
\end{aligned}
\end{equation}
where $\mu$ is an eigenvalue of $-\D_S$. 

This equation can only be defined in $(-r_0,r_0)$, \resp in $\R[]$,  if
\begin{equation}\lae{6.7}
\mu=a^{-1}_0\tfrac{p(p-2)}4 
\end{equation}
is an eigenvalue of $-\D_S$.

Thus, we have to assume  $m\ge 2$ and  in addition that $\mu$ in \re{6.7} is an eigenvalue.

Let $\h\in E_\mu$ be an arbitrary eigenvector of unit length. 

A wave function $\psi=\psi(r,y^A)$ then satisfies
\begin{equation}
\hat H\psi=0,
\end{equation}
if 
\begin{equation}
\psi=u\h,\q\h\in E_\mu,
\end{equation}
and $u=u(r)$ is a solution of
\begin{equation}\lae{6.10}
\Ddot u 
+4(\bar V+\bar\Lam)r^{2(n-1)}u-4\tilde \ka r^{2(n-2)}u=0. 
\end{equation}

Let $u$ be a real valued function defined in $(0,r_0)$, then we look at the eigenvalue problem \re{6.10} with boundary conditions
\begin{equation}\lae{6.11}
u(0)=u(r_0)=0,
\end{equation}
where $\bar\Lam$ is supposed to be an eigenvalue, \cf \rs{5},\footnote{The results in \rs 5 are also valid for the present form of the Hamiltonian, though the phrasing in the proofs would have to be modified occasionally.}  and we conclude as before that the eigenvalue problem has countably many solutions $(\bar\Lam_i,u_i)$ such that
\begin{equation}
\bar\Lam_i<\bar\Lam_{i+1}\qq\A\,i\in\N,
\end{equation}
\begin{equation}
\lim_i\bar\Lam_i=\un,
\end{equation}
and that the eigenspaces are one dimensional, and where $u_0$ doesn't vanish in $(0,r_0)$.

The same result is valid in $(-r_0,0)$, where $r$ in \re{6.10} has to be replaced by $(-r)$.
\bd
Let $\bar\Lam_i$ be an eigenvalue of the operator in \re{6.10}. A function $\tilde u\in C^\un((-r_0,r_0))$  is said to represent a smooth transition from big crunch to big bang, if it satisfies the equation \re{6.10} in $(-r_0,r_0)$ and the restrictions 
\begin{equation} 
\fv{\tilde u}{(-r_0,0)}\q\text{\resp}\q \fv{\tilde u}{(0,r_0)}
\end{equation}
belong to the respective eigenspaces $E_{\bar\Lam_i}$ such that $\dot{\tilde u}(0)\ne 0$.
\ed
\bt
Let $(\bar\Lam_i,u_i)$ be a solution of the eigenvalue problem \re{6.10}, \re{6.11}, and define
\begin{equation}
\tilde u(r)=
\begin{cases}
\hp{-}u(r),\;\hp{-} r>0,\\
-u(-r),\;r<0,
\end{cases}
\end{equation}
then $\tilde u$ is a $C^\un$-transition from big crunch to big bang for arbitrary $n\ge 3$. 
\et
\bp
$\tilde u$ evidently solves the equation in $(-r_0,0)$ \resp $(0,r_0)$. We shall show 
\begin{equation}
\tilde u\in C^2((-r_0,r_0)),
\end{equation}
from which the additional claims easily follow. 

$\tilde u$ is certainly of class $C^1$. It is also of class $C^2$, since
\begin{equation}
\Ddot u(0)=0,
\end{equation}
in view of the equation. Hence $\tilde u$ solves the equation in $(-r_0,r_0)$, and the additional claims follow from well-known regularity theorems. In fact $\tilde u$ is even real analytic.
\ep

\br
The theorem is also valid in the unbounded case.
\er

\section{The implicitly defined Hamiltonians and the Schr\"odinger equation} \las{8}
Let $H$ be the Hamilton operator in \fre{1.11}, then the equation
\begin{equation}\lae{7.1.1}
H\psi=0
\end{equation}
defines two Hamiltonians $H_i$, $i=1,2$, where 
\begin{equation}
\begin{aligned}
H_1\psi&=-r^{-1}(r\dot\psi)'-r^{-2}\big(-\tfrac 4{n^2}a_0\D\psi -\tfrac{(m-1)^2}4\psi\big) \\[\cma]
&\q\;-\tfrac{16}{n^2}\bar V r^{2}\psi +\tfrac{16}{n^2}\tilde\ka r^\frac{2(n-2)}{n}\psi
\end{aligned}
\end{equation}
and
\begin{equation}
\begin{aligned}
H_2\psi&=-r^{-1}(r\dot\psi)'-r^{-2}\big(-\tfrac 4{n^2}a_0\D\psi -\tfrac{(m-1)^2}4\psi\big) \\[\cma]
&\q\;+\tfrac{16}{n^2}r^{2}\psi.
\end{aligned}
\end{equation}

The $H_i$ are defined in dense subspaces of $L^2(I_i,1)\otimes\mc H_0$, where
\begin{equation}
I_1=(0,r_0)\q\wed\q I_2=(0,\un),
\end{equation}
and where we observe that
\begin{equation}\lae{7.5}
\begin{aligned}
\mc H(I_i,1)=L^2(I_i,1)\otimes \mc H_0=\bigoplus_{k=0}^{k_0}L^2(I_i,1)\otimes E_{\mu_k},
\end{aligned}
\end{equation}
where $E_{\mu_k}$ is an eigenspace of $-\D$ such that the eigenvalues $\mu_k$ satisfy the condition \fre{1.25.2}.

Thus, any $\psi\in \mc H(I_i,1)$ can be expressed as
\begin{equation}\lae{7.6}
\psi=\sum_{k=0}^{k_0}u_k\h_k,
\end{equation}
where $\h_k\in E_{\mu_k}$ and $u_k\in L^2(I_i,1)$.
\bd\lad{7.1}
 Let $\mc H_i(I_i)$ be the completions of $C^\un_c(I_i)\otimes\mc H_0$ with respect to the scalar product norms
\begin{equation}
\norm\psi^2=\sum_{k=0}^{k_0}\int_{I_1}\{r\abs{\dot u_k}^2-r^{-1}\bar\mu_k\abs{u_k}^2\}\norm{\h_k}^2
\end{equation}
in case $i=1$, and
\begin{equation}
\norm\psi^2=\sum_{k=0}^{k_0}\int_{I_1}\{r\abs{\dot u_k}^2-r^{-1}\bar\mu_k\abs{u_k}^2+r^3\abs{u_k}^2\}\norm{\h_k}^2 
\end{equation}
for $i=2$, where we used the expression \re{7.6} for $\psi$, and where $\norm{\h_k}$ is the norm in $L^2(S)$, and where we stipulate that in case of $H_2$ the conditions \fre{1.36.2} have to be satisfied with $\bar V=0$.

Both spaces can be compactly embedded in $\mc H(I_i,1)$ respectively. 
\ed
\bt\lat{7.2.1}
The eigenvalue problems in $\mc H_i(I_i)$, $i=1,2$,
\begin{equation}\lae{7.11.1}
H_1\psi=\bar\Lam \tfrac{16}{n^2}r^{2}\psi
\end{equation}
\resp
\begin{equation}
H_2\psi=-\lam \tilde\ka \tfrac{16}{n^2}r^\frac{2(n-2)}{n}\psi
\end{equation}
have countably many solutions $(\bar\Lam_j,\psi_j)$ \resp $(\lam_j,\psi_j)$ such that
\begin{equation}
\bar\Lam_j\le \bar\Lam_{j+1}\q\wed\q \lim_j\bar\Lam_j=\un
\end{equation}
and
\begin{equation}
\lam_j\le\lam_{j+1}\q\wed\q \lim_j\lam_j=\un.
\end{equation}
The respective eigenvectors $(\psi_j)$ are complete in $\mc H_i(I_i)$ as well as in $\mc H(I_i,1)$.
\et
\bp
We only consider the case $H_1$, since the arguments for $H_2$ are identical.

 \cq{Existence of $(\bar\Lam_j,\psi_j)$}\q Using the general variational principle to solve abstract eigenvalue problems that we cited in \rs{5}, we could prove the claims of the theorem. On the other hand, given the orthogonal decomposition of $\mc H(I_1,1)$ in \re{7.5} we deduce that the finite family of eigenvalues and eigenfunctions, the existence of which we proved in \frt{1.5} for each $E_{\mu_k}$, $0\le k\le k_0$, are a partition of the solutions of the present eigenvalue problem for $H_1$.

Notice that the eigenvalue equations are to be understood in the distributional sense, \cf  \rd{7.3} below.
\ep

If we want to define selfadjoint Hamiltonians, then slightly altered Hamiltonians have to be considered.

\bd\lad{7.3}
Let $p_i$, $i=1,2$, be equal to
\begin{equation}
p_1=2\q\wed\q p_2=\tfrac{2(n-2)}n
\end{equation}
and define the Hamiltonians $\hat H_i$ by
\begin{equation}
\hat H_i=r^{-p_i} H_i.
\end{equation}

The previously defined spaces $\mc H_i(I_i)$ can also be compactly embedded in 
\begin{equation}
\mc H(I_i,p_i+1)=L^2(I_i,p_i+1)\otimes \mc H_0,
\end{equation}
and we define the domains of $\hat H_i$ by
\begin{equation}
D(\hat H_i)=\set{\psi\in\mc H_i(I_i)}{\hat H_i\psi \in \mc H(I_i,p_i+1)},
\end{equation}
where the last condition is to be understood in the distributional sense, i.e., $\hat H_i\psi=\chi$ means
\begin{equation}
\spd{\psi}{\hat H_i\f}=\spd\chi\f\qq\A\,\f\in C^\un_c(I_i)\otimes\mc H_0.
\end{equation}
\ed

We can now prove: 
\bt\lat{7.4}
The Hamiltonians $\hat H_i$ are selfadjoint in $\mc H(I_i,p_i+1)$ with a pure point spectrum and the eigenvalue problems
\begin{equation}
\hat H_1\psi=\bar\Lam \tfrac{16}{n^2}\psi
\end{equation}
\resp
\begin{equation}
\hat H_2\psi=-\lam \tilde\ka \tfrac{16}{n^2}\psi
\end{equation}
have countably many solutions $(\bar\Lam_j,\psi_j)$ \resp $(\lam_j,\psi_j)$ such that
\begin{equation}
\bar\Lam_j\le \bar\Lam_{j+1}\q\wed\q \lim_j\bar\Lam_j=\un
\end{equation}
and
\begin{equation}
\lam_j\le\lam_{j+1}\q\wed\q \lim_j\lam_j=\un.
\end{equation}
The respective eigenvectors $(\psi_j)$ are complete in $\mc H_i(I_i)$ as well as in $\mc H(I_i,p_i+1)$.
\et
\bp
It suffices to prove the selfadjointness of the operators, since the statements about the eigenvalues can be proved by the same arguments as in case of \rt{7.2.1}.

Again we only give a proof for the Hamiltonian $\hat H_1$.

\cvm
(i) \cq{$\hat H_1$ is closed.}\q Let $\psi_k\in D(\hat H_1)$ be a sequence such that
\begin{equation}
\psi_k\ra \psi\q\wed\q \hat H_1\psi_k\ra\chi.
\end{equation}
We have to show that $\psi\in D(\hat H_1)$ and
\begin{equation}
\hat H_1\psi=\chi
\end{equation}
in the distributional sense.

First we observe that the $\psi_k$ are bounded in $\mc H_1(I_1)$. Hence, a subsequence, not relabelled, converges weakly in $\mc H_1(I_1)$ to some element $\tilde\psi$, which has to be identical to $\psi$, since
\begin{equation}
\mc H_1(I_1)\hra \mc H (I_1,p_1+1).
\end{equation}

Next let $\f\in C^\un_c(I_1)\otimes\mc H_0$, then
\begin{equation}
\spd{\hat H_1\psi_k}\f=\spd{\psi_k}{\hat H_1\f}
\end{equation}
and
\begin{equation}
\spd{\hat H_1\psi_k}\f\ra \spd\chi\f
\end{equation}
while
\begin{equation}
\spd{\psi_k}{\hat H_1\f}\ra \spd\psi{\hat H_1\f}=\spd{\hat H_1\psi}\f.
\end{equation}
Thus we conclude $\psi\in D(\hat H_1)$ and $\hat H_1\psi=\chi$.

\cvm
(ii) \cq{$\hat H_1$ is selfadjoint.}\q  Let $c>0$ be such that
\begin{equation}\lae{7.21.2}
\spd{\hat H_1\psi}\psi+c\spd\psi\psi\ge c_0\norm\psi^2
\end{equation}
for some $c_0>0$. Then we shall prove that
\begin{equation}
R(\hat H_1+cI)=\mc H(I_1,p_1+1)
\end{equation}
which will imply the selfadjointness of $\hat H_1$.

First we note that $R(\hat H_1+cI)$ is closed. Indeed, let
\begin{equation}
\hat H_1\psi_k+c\psi_k
\end{equation}
be a Cauchy sequence, then $\norm{\psi_k}\le\const$, because of \re{7.21.2}, and thus, the $\psi_k$ are also a Cauchy sequence in $\mc H(I_1,p_1+1)$; hence, the result in view of (i).

Second, we observe that $R(\hat H_1+cI)$ is dense in $\mc H(I_1,p_1+1)$, since it contains the eigenvectors $(\psi_j)$ which are complete as we already proved.
\ep
For the Hamiltonians $\hat H_i$ we can solve  the corresponding Schr\"odinger equations. Again we shall only consider the Hamiltonian $\hat H_1$ in detail.

Any eigenvector $\psi_0$ of $\hat H_1$ defines a solution $\psi=\psi(t,\cdot)$ of the Schr\"odinger equation
\begin{equation}\lae{8.31}
\pde \psi t=-i \hat H_1\psi
\end{equation}
in $\mc H(I_1,p_1+1)$ by defining
\begin{equation}\lae{8.32}
\psi=e^{-it\hat H_1}\psi_0=e^{-it\bar\Lam \frac{16}{n^2}}\psi_0.
\end{equation}

Even if $\psi_0$ is not an eigenfunction, the right-hand side of the first equation in \re{8.32} is a solution of the Schr\"odinger equation, if $\psi_0\in D(\hat H_1)$.  However, we emphasize that only eigenstates or superpositions of eigenstates of $\hat H_1$ are actual solutions of the original problem  \re{7.1.1}, \cf  however \frr{10.4}, where we argue that the \cq{admissible} wave functions should be those belonging to $\mc H_1(I_1)$.

\section{The problem of time}\las{9}
Since the wave functions in \fre{1.11} only depend on $r$ and $\f^A$ and not explicitly on any known time variable, it has been argued that the quantum cosmological universe is stationary without any time orientation or dynamical development. However, in classical Friedman universes the variable $r$ is a time function, at least in the weak sense, i.e., the level hypersurfaces $M_\rho$=$\{r=\rho=\const\}$ are spacelike hypersurfaces and apart from the case when $\rho$ is a \tit{singular} value,\footnote{Notice that by Sard's theorem the set of singular values has measure zero.} $\grad r=0$, $M_\rho$ is then a maximal hypersurface, $r$ is also a time function in the usual sense, i.e., its gradient is timelike.

Hence, it seems natural to use $r$, or any invertible function of $r$, as a measure for time in quantum cosmological Friedman models. Misner \cite{misner:time} suggested to use
\begin{equation}
\Om=-\log \abs{M_r}=- n\log r,
\end{equation}
as a time function, where $\abs{M_r}$ is the volume of $M_r$ and the volume of $M_1$ has been set to $1$. 

The distinction between a big bang and a big crunch can then be defined by the sign of $r$: If the singularity is specified by $r=0$, then it will lie in the past, if $r>0$, and in the future, if $r<0$. The time direction in quantum cosmological Friedman models can be defined by increasing or decreasing values of $r$; switching from $r$ to $-r$ will change this time direction. Hence there will always exist two quantum cosmological Friedman models with opposite time direction, one with $r>0$ and the other with $r<0$, but in both cases the singularity is supposed to be in the past. Another way to express opposite time directions would be to choose $r>0$ in both cases, but referring to the singularity $r=0$ as a past \resp future event.

Using the selfadjoint Hamiltonians in \rs{8} to formulate the Wheeler-DeWitt equation in the bounded and unbounded models, we have an alternate method to define time  and a dynamical development of the wave functions. It is well-known in quantum theory that a selfadjoint operator $H$ is the generator of a continuous one-parameter group of unitary transformations $U(t)$,
\begin{equation}
U(t)=e^{-itH},
\end{equation}
such that the dynamical development of a wave function $\psi_0$ is given by the solution $\psi(t)$ of the Schr\"odinger equation \fre{8.31}  with initial value $\psi_0$, \cf \cite[Chapter 2.3]{mackey:book}. Since only solutions of the Schr\"odinger equation that are also solutions of the Wheeler-DeWitt equation are physically reasonable, the initial value has to be either a pure eigenfunction or a finite superposition of eigenfunctions. In the latter case this will lead to a non-trivial dynamic development of the initial state; a pure eigenstate, of course, remains stationary.

\section{Comparison with the classical Friedman solutions}\las{10} 

In this section we want to compare the solutions of the classical Friedman equation corresponding to the functional \fre{1.10} with the quantum solutions, where we are especially interested in the range of the scale factor $r$.

We shall prove that in both approaches---classical and quantum theoreti\-cal---there is a big bang singularity but that the range of $r$ is just opposite of each, i.e., when $r$ is classically bounded, it will be unbounded in the quantum cosmological setting and vice versa, if we consider only fixed values for the cosmological constant. 

However, we should bear in mind that in the quantum cosmological setting we always have to consider a sequence of cosmological constants, which tend to infinity in the bounded model and to zero in the unbounded one. If we include these sequences $(\Lam_i)$ into our comparison then it turns out that the only discrepancy occurs in case of the bounded quantum cosmological model, its classical counterparts will all have unbounded radii, \cf \rt{10.3} below, while the classical counterparts of the unbounded quantum model have a bounded radius $r_i$ for each $\Lam_i<0$, but these radii tend to infinity, \cf \rt{10.2}. 

A classical solution has to satisfy the Friedman equation
\begin{equation}
G_{00}+\Lambda \bar g_{00}=\frac{\al_M}2 T_{00},
\end{equation}
which in our case looks like
\begin{equation}
\tfrac12 n(n-1)\tilde\ka+\tfrac12 n(n-1) \abs{f'}^2-\Lam e^{2f}=\frac{\al_M}2\rho e^{2f},
\end{equation}
where we used that the metric is expressed in the conformal time gauge $\tau$ as in \fre{3.9.7} denoting\begin{equation}
f=\log r\q\wed\q f'=\df f\tau,
\end{equation}
and
\begin{equation}
\rho=\tfrac12 \norm{\dot\f}^2e^{-2f}+V;
\end{equation}
notice that a dot or prime have the same meaning.

Assuming without loss of generality $V=0$, since $V=\const$ acts like a cosmological constant, we conclude
\begin{equation}
\rho r^{2n}=\rho e^{2nf}=c_0=\const>0,
\end{equation}
\cf \cite[Lemma 4.4]{cg:brane2}, and we deduce further that the Friedman equation is equivalent to
\begin{equation}\lae{10.6}
\begin{aligned}
\abs{f'}^2=\frac{\al_M}{n(n-1)} c_0 e^{-2(n-1)f}+\frac{2\Lam}{n(n-1)}e^{2f}-\tilde\ka\equiv D=D(f).
\end{aligned}
\end{equation}

The qualitative behaviour of a solution $f$ depends on the fact if $D$ vanishes for some values of $f$ or if $D>0$ everywhere.

From the definition of $D$ we immediately conclude
\begin{equation}
\Lam<0\im D^{-1}(0)\ne \eS,
\end{equation}
while in case $\Lam \ge 0$ there holds:
\bl
Let $\Lam \ge 0$, then $D>0$ if $\tilde\ka\le 0$, and
\begin{equation}\lae{10.8}
\al_M^{\frac1n}c_0^\frac1n\Lam^\frac{n-1}n>c_1\tilde\ka,\q c_1=c_1(n),
\end{equation}
implies $D>0$ if $\tilde\ka>0$.
\el

\bp
Obviously, we only have to consider the case $\tilde\ka>0$ and $\Lam>0$. Then $D$ attains its infimum for finite $f$, or equivalently, setting  
\begin{equation}
r=e^{2f},
\end{equation}
for a unique $r>0$ satisfying
\begin{equation}
\df Dr=-\frac{\al_M}n c_0 r^{-n}+\frac{2\Lam}{n(n-1)}=0,
\end{equation}
hence for
\begin{equation}
r=\big(\frac{(n-1)\al_Mc_0}{2\Lam}\big)^\frac1n.
\end{equation}
Evaluating $D$ for that particular $r$ we obtain
\begin{equation}\lae{10.12}
D=\al_M^\frac1nc_0^\frac1n\Lam^\frac{n-1}n2^\frac{n-1}n (n-1)^{-\frac{2n-1}n}-\tilde\ka,
\end{equation}
hence the result.
\ep

In order to solve the Friedman equation we first switch to the eigen time gauge $t$ such that
\begin{equation}
dt=e^f d\tau.
\end{equation}
Defining then
\begin{equation}
f'=\df ft
\end{equation}
the Friedman equation takes the form
\begin{equation}
\begin{aligned}
\abs{f'}^2=De^{-2f}=\ga_1 e^{-2nf}+\ga_2\Lam-\tilde\ka e^{-2f}\equiv\tilde D=\tilde D(f)
\end{aligned}
\end{equation}
with positive constants $\ga_i$, $i=1,2$.

We first consider the case $\Lam<0$.

\bt\lat{10.2} 
Let $\Lam<0$ and $\tilde\ka\le 0$, then there exists $a>0$ and $f\in C^\un (I)$, where $I=(-a,a)$, solving the initial value problem
\begin{equation}\lae{10.16}
\begin{aligned}
f'&=\begin{cases}
\hp{-}\tilde D^\frac12, \;t<0,\\
-\tilde D^\frac12, \,t>0, 
\end{cases}\\[\cma]
f(0)&=f_0,
\end{aligned}
\end{equation}
where $\tilde D(f_0)=0$.

$f$ is an even function
\begin{equation}
f(t)=f(-t)
\end{equation}
and
\begin{equation}\lae{10.18}
\lim_{t\ra -a}f(t)=\lim_{t\ra a}f(t)=-\un,
\end{equation}
i.e., the Friedman universe has a big bang as well as a big crunch singularity and is time symmetric with respect to the unique maximal hypersurface $M_0=\{t=0\}$.

Moreover, if $\tilde\ka<0$, then the estimates
\begin{equation}\lae{10.19.7}
-\Lam e^{2f_0}\ge -\tfrac{n(n-1)}2\tilde\ka
\end{equation}
and
\begin{equation}\lae{10.20.7}
a\ge c_1{\abs\Lam}^{-\frac12}-c_2\abs\Lam^{-\frac1{2n}},
\end{equation}
where $c_i=c_i(n,\tilde\ka,c_0)>0$, $i=1,2$, are valid, i.e., the maximal radius as well as the Lorentzian diameter of the universe will tend to infinity if $\Lam$ tends to zero.
\et

\bp
Define
\begin{equation}
\tilde D'=\df {\tilde D}f
\end{equation}
and assume for the moment that the initial value problem \re{10.16} had already a smooth solution. Differentiating the differential equation in \re{10.16} with respect to $t$ for $t<0$ we deduce
\begin{equation}\lae{10.20}
\begin{aligned}
f''=\tfrac12 \tilde D^{-\frac12}\tilde D'f'=\tfrac12 \tilde D'=-n\ga_1 e^{-2nf}+\tilde\ka e^{-2f}.
\end{aligned}
\end{equation}

The right-hand side is a smooth function, hence the initial value problem 
\begin{equation}
\begin{aligned}
f''&=\tfrac12\tilde D',\\
\{f'(0),f(0)\}&=\{0,f_0\},
\end{aligned}
\end{equation}
has a smooth solution on a maximal open interval $I$.

Because of the uniqueness of the solution we immediately conclude that $f$ is even.

Moreover, $f$ also solves \re{10.16}. To prove this claim it suffices to show that
\begin{equation}
\abs{f'}^2=\tilde D.
\end{equation}
Let
\begin{equation}
w=\abs{f'}^2-\tilde D,
\end{equation}
then
\begin{equation}
\begin{aligned}
w'=2f'f''-\tilde D'f'
=f'(2f''-\tilde D')=0,
\end{aligned}
\end{equation}
hence
\begin{equation}
w\equiv\const=w(0)=0.
\end{equation}

Since $f$ is even, the maximal interval is of the form $I=(-a,a)$, and it remains to prove $a<\un$ and the other relations. 

\cvm
\cq{$a<\un$}\q Consider the interval $0<t<a$. Then we conclude that for any $0<t_0<a$ there exists $0<\e_0<\ga_1$ such that
\begin{equation}
(\ga_1-\e_0^2)e^{-2nf(t_0)}+\ga_2\Lam-\tilde\ka e^{-2f(t_0)}>0.
\end{equation}
Since $f'<0$ and $\tilde\ka\le 0$ we deduce further that
\begin{equation}
\tilde D(f(t))-\e_0^2e^{-2nf(t)}\ge \tilde D(f(t_0))-\e_0^2e^{-2nf(t_0)}>0,\q\A\,t>t_0
\end{equation}
and therefore
\begin{equation}
f'\le -\e_0e^{-nf}\q\A\,t\ge t_0.
\end{equation}

Multiplying this inequality by $e^{nf}$ and integrating we then obtain
\begin{equation}
\tfrac1n\{e^{nf(t)}-e^{nf(t_0)}\}\le -\e_0(t-t_0)
\end{equation}
and infer $a<\un$.

\cvm
\cq{\re{10.18} is valid.}\q Since $I$ is a finite interval and maximal  we conclude that the right-hand side of \re{10.20} must become unbounded, if $t$ tends to $a$ or $-a$, hence the result.

\cvm
\cq{\re{10.19.7} is valid.}\q This claim follows immediately from \re{10.6} and the defining property of $f_0$, namely, $D(f_0)=0$.

\cvm
\cq{\re{10.20.7} is valid.}\q Since
\begin{equation}
D+\tilde\ka=
\begin{cases}
\tilde\ka,&t=0,\\
\un,&t=a,
\end{cases}
\end{equation}
and $\tilde\ka<0$, there exists $0<t_0<a$ be such that
\begin{equation}\lae{10.33.7}
\ga_1 e^{-2nf(t_0)}=-\ga_2\Lam.
\end{equation}

Moreover, $\tilde D+\tilde\ka e^{-2f}$ is monotonically increasing with respect to $t$, hence we infer
\begin{equation}
\tilde D\le -\tilde\ka e^{-2f}\qq\A\,0<t<t_0,
\end{equation}
and therefore
\begin{equation}
f'\ge -(-\tilde\ka)^\frac12 e^{-f}
\end{equation}
from which we deduce by integrating
\begin{equation}
e^{f(t_0)}-e^{f_0}\ge -(-\tilde\ka)^\frac12 t_0,
\end{equation}
or equivalently
\begin{equation}
(-\tilde\ka)^\frac12 t_0\ge e^{f_0}-e^{f(t_0)}.
\end{equation}

The result then follows by taking the relations \re{10.19.7} and \re{10.33.7} into account.
\ep

Next, let us consider the case $\Lam>0$.

\bt\lat{10.3}
Let $\Lam>0$ be so large that $\tilde D>0$ everywhere. Then the initial value problem
\begin{equation}\lae{10.30}
\begin{aligned}
f'&=\tilde D^\frac12=(\ga_1 e^{-2nf}+\ga_2\Lam-\tilde\ka e^{-2f})^\frac12,\\
f(0)&=f_0,
\end{aligned}
\end{equation}
 has a smooth solution on  a maximal open interval $(a,\un)$ such that
 \begin{equation}
\lim_{t\ra a}f(t)=-\un
\end{equation}
and
\begin{equation}\lae{10.32}
\lim_{t\ra \un}f(t)=\un,
\end{equation}
i.e., the universe has a big bang but no big crunch singularity and its Lorentzian diameter is unbounded as well as its radial diameter.
\et

\bp
Let $(a,b)$ be the maximal open interval in which the solution of \re{10.30} exists. Looking at the minimum value of $D$ in equation \re{10.12} we deduce that there exists $\e_0>0$ such that
\begin{equation}\lae{10.33}
\begin{aligned}
f'=\tilde D^\frac12=D^\frac12 e^{-f}\ge \e_0e^{-nf},
\end{aligned}
\end{equation}
and hence
\begin{equation}
\tfrac1n(e^{nf(0)}-e^{nf(t)})\ge -\e_0t\q\A\, a<t<0,
\end{equation}
hence $a\in\R[]$.

Obviously, there is a big bang singularity in $t=a$.

\cvm
\cq{$b=\un$}\q  For any $a<t_0<b$ we infer
\begin{equation}
f'\le\const\q\A\,t_0<t<b,
\end{equation}
hence $b=\un$, since it is supposed to be is maximal.

\cvm
\cq{\re{10.32}}\q Using \re{10.33} we conclude
\begin{equation}
\tfrac1n(e^{nf(t)}-e^{nf(0)})\ge\e_0t\q\A\, 0<t<\un,
\end{equation}
 completing the proof.
\ep
\section{Concluding remarks} 
\br\lar{10.4}
In the previous sections we have shown that the Wheeler-DeWitt equation  shouldn't be considered as a differential equation for a hyperbolic differential operator with fixed coefficients but as an implicit eigenvalue equation where the cosmological constant, or an algebraic manipulation of it, is supposed to assume the respective eigenvalues. Any eigenfunction or superposition of eigenfunctions is then also a solution of the Wheeler-DeWitt equation. Since the eigenfunctions are complete in the respective Hilbert spaces $\mc H_i(I_i)$, \cf \frd{7.1}, and also in the larger spaces $\mc H(I_i,p_i+1)$, \cf \frd{7.3}, it is justified to say that any vector $\psi\in \mc H_i(I_i)$ is an \tit{admissible} wave function for the corresponding model. We don't apply this attribute to elements merely  belonging to  the larger spaces, since only the functions in $\mc H_i(I_i)$ have finite expectation values with respect to the implicitly defined Hamiltonians.
\er
\br\lar{10.5}
Finding a spectral resolution for the operator corresponding to the  Wheeler-DeWitt equation we had to prove that the implicitly defined Hamiltonians are not only selfadjoint but also have a pure point spectrum for otherwise the Wheeler-DeWitt equation wouldn't be satisfied. Under these circumstances only bounded models are possible for positive cosmological constants.

Negative cosmological constants corresponding to infinitely many eigenvalues, i.e., to large quantum numbers, are only possible if we consider an unbounded model with $\tilde\ka<0$, because of the transformation in \fre{6.40.7} requiring $0<r<\un$.
\er

\providecommand{\bysame}{\leavevmode\hbox to3em{\hrulefill}\thinspace}
\providecommand{\href}[2]{#2}


\end{document}